\documentclass[11pt]{article}
\usepackage{jcappub}
\usepackage{mathtools}
\usepackage{bm}
\usepackage{dsfont}
\usepackage{color}
\usepackage{array}
\usepackage{graphicx}
\usepackage{subcaption}
\usepackage{soul}
\usepackage{multirow}
\usepackage{multicol}
\usepackage{float}
\usepackage{hhline}
\usepackage[dvipsnames]{xcolor}
\usepackage[normalem]{ulem}
\usepackage{mathrsfs}
\usepackage{wasysym}
\usepackage[mathscr]{euscript}
\usepackage{cases}

\title{Axial Quasi-normal Modes of Boson-Fermion Stars}

\author[a]{Hamza Boumaza,}
\author[b]{Boris Betancourt Kamenetskaia}

\affiliation[a]{Laboratoire de Physique des Particules et Physique Statistique (LPPPS),
 Ecole Normale Supérieure-Kouba, B.P. 92, Vieux Kouba, 16050 Algiers, Algeria}
 \affiliation[b]{Cosmology, Gravity, and Astroparticle Physics Group, Center for Theoretical Physics of the Universe,
Institute for Basic Science (IBS), Daejeon, 34126, Korea}

\emailAdd{hamza.boumaza@g.ens-kouba.dz}
\emailAdd{laybors@ibs.re.kr}

\abstract{We study axial quasi-normal modes of boson-fermion stars composed of ordinary nuclear matter and a self-interacting bosonic dark matter component. The equilibrium configurations are obtained by solving the coupled two-fluid Tolman–Oppenheimer–Volkoff equations, where the neutron sector is modeled with several realistic equations of state and the bosonic sector is described by a repulsively self-interacting complex scalar field in the strong-coupling regime. We analyze linear axial perturbations governed by a Regge–Wheeler type equation whose effective potential reflects the combined matter distribution. Using a continued-fraction method, we compute the complex eigenfrequencies of the fundamental and overtone $w$ modes. We obtain the quasi-normal mode spectrum and investigate its dependence on the dark matter particle mass, self-coupling, and the central densities of both fluids for several realistic neutron star equations of state. We find that increasing the dark matter fraction shifts the oscillation frequencies and damping times. It can also reorder the mode hierarchy through crossings, and it drives a continuous transition from neutron star-like to boson star-like ringdown behavior. Our results demonstrate that the ringdown gravitational-wave signal from post-merger compact objects could encode clear imprints of a dark matter component, offering a new probe of the dark sector with future gravitational-wave observatories.}

\begin{document}
\maketitle
\flushbottom

\section{Introduction}

Despite robust evidence for the existence of dark matter (DM) on galactic, cluster, and cosmological scales, its fundamental nature remains elusive (for reviews, see~\cite{Jungman:1995df,Bertone:2004pz,Bergstrom:2000pn,Feng:2010gw}). Key properties of the DM particle, namely its mass, spin, self-interactions, and possible couplings to the visible sector, remain largely unconstrained. In many well-studied scenarios, the DM is taken to be a stable, self-conjugate particle with weak-scale interactions, a picture originally motivated by the lightest neutralino in supersymmetry~\cite{Ellis:1983ew}. From a phenomenological viewpoint, however, these assumptions need not hold. In recent years, a wide variety of models have relaxed them, leading to a much richer landscape of DM candidates and astrophysical signatures.

A particularly compelling possibility is that DM is asymmetric, hence carrying a conserved “dark charge” akin to baryon number, with an excess of particles over antiparticles established in the early Universe~\cite{Barr:1990ca,Gudnason:2006yj} (for reviews, see \cite{Zurek:2013wia,Petraki:2013wwa}). In this scenario, the residual DM population is stable and cannot self-annihilate, mirroring the baryon asymmetry of ordinary matter. If DM also possesses strong self-interactions, a fraction of the asymmetric component can gravitationally collapse, forming stable compact objects commonly referred to as dark stars. The idea that such gravitationally bound structures could exist dates back to Kaup~\cite{Kaup:1968zz}, who studied a system of non-interacting complex scalar fields supported against collapse by the uncertainty principle. Later, Colpi, Shapiro, and Wasserman~\cite{Colpi:1986ye} extended this analysis to include repulsive self-interactions among the scalar particles, demonstrating that self-interacting bosonic configurations, often called boson stars (BSs), can reach macroscopic masses and sizes, with distinct equilibrium properties controlled by the strength of the self-coupling.

These dark stars could contribute to the DM budget of galaxies and potentially be detected through a variety of gravitational probes. Observational constraints from stellar microlensing~\cite{Macho:2000nvd,EROS-2:2006ryy,Niikura:2019kqi}, supernova magnification~\cite{Zumalacarregui:2017qqd}, gravitational waves from dark compact objects~\cite{Maselli:2017vfi,Kavanagh:2018ggo,LIGOScientific:2019kan,Chen:2019irf,BetancourtKamenetskaia:2026cyf}, and dynamical friction in dwarf galaxies~\cite{Brandt:2016aco} collectively suggest that at most a few percent of the galactic DM can reside in compact dark stars, with the precise limit depending on their mass spectrum (cf.~\cite{Green:2020jor}). Beyond pure gravitational signatures, dark stars could emit electromagnetic radiation through kinetic mixing with photons~\cite{Maselli:2019ubs}, or, if the stabilizing dark charge is slightly broken, via annihilation or decay of DM particles inside the star~\cite{Kamenetskaia:2022lbf}.

Quasi-normal modes (QNMs) are the characteristic damped oscillations of compact objects and provide a powerful probe of the internal structure of relativistic stars through gravitational-wave observations. For neutron stars (NSs), the study of non-radial oscillations was initiated in Refs.~\cite{thorne1969nonradial,campolattaro1970nonradial,thorne1967non}, where the relativistic perturbation framework was established. Later, Chandrasekhar and Ferrari~\cite{Chandrasekhar:1991fi} derived the equations governing the non-radial oscillations of a static, spherically symmetric matter distribution and showed that the problem reduces to the determination of complex eigenfrequencies, namely the QNMs. They also demonstrated that the perturbation equations decouple into two independent sectors: odd-parity perturbation, which contains a single dynamical degree of freedom associated with spacetime dynamics, and even-parity perturbations, which involve two coupled degrees of freedom related to both fluid motion and gravitational dynamics.

 The odd-parity (axial) sector is particularly interesting because it is governed primarily by spacetime dynamics and only indirectly by the stellar matter through the equilibrium background. Therefore, axial modes provide one of the cleanest theoretical probes of how changes in the internal structure of compact stars, such as the addition of DM, modify the gravitational-wave spectrum. While axial modes typically lie at high frequencies ($\gtrsim$ several kHz) and are weakly excited in astrophysical mergers compared to fluid-driven even-parity modes, they are nonetheless an essential part of the full QNM spectrum. Their simplicity makes them ideal for systematic studies of the relationship between the stellar interior and the gravitational response. Thus, while axial modes are not expected to dominate current ringdown observations, they provide a theoretically clean diagnostic of the spacetime structure and are an essential ingredient in a complete description of compact-star oscillations.\\
 
The axial $w$-mode spectrum of  relativistic stars has been  investigated over the past several decades. The authors of Refs.~\cite{Kokkotas:1992lxs,Kokkotas:1994an,Kokkotas:2002sf} presented the existence of spacetime $w$-modes and analyzed their properties, showing that these modes are mainly associated with perturbations of the spacetime curvature rather than with fluid motion. For more details on the classification of the different oscillation modes of  compact stars, we refer the reader to the reviews 
\cite{Kokkotas:1999bd,Barack:2018yly}. The spectrum of NSs for different realistic equations of state is investigated 
in Ref.~\cite{Andersson:1997rn}, discussing their relevance for gravitational-wave emission. The axial $w$-mode spectrum of NSs with exotic matter has also been extensively studied in the literature. Blázquez-Salcedo \textit{et al.}~\cite{Blazquez-Salcedo:2012hdg,Blazquez-Salcedo:2013jka} analyzed the effect of hyperons and quarks on the axial $w$-modes using a large set of realistic equations of state, finding that the spacetime modes can help discriminate between purely hadronic stars and stars containing exotic matter for sufficiently compact configurations. Chatterjee and Bandyopadhyay~\cite{2009PhRvD..80b3011C} performed a similar analysis which included antikaon condensates and showed that the fundamental axial $w$-mode shifts to higher frequencies in the presence of condensates. More recently,~\cite{Dey:2025wcs} studied the $w$-modes of a compact object composed of quarkyonic matter and fermionic dark matter. \\

In this work, we study NSs whose interiors contain ordinary baryonic matter together with an additional DM component, thereby requiring a two-fluid description. The presence of this extra fluid can significantly modify the stellar structure, affecting the mass-radius (MR) relation~\cite{Leung:2011zz,Xiang:2013xwa}, tidal deformability~\cite{Ellis:2018bkr}, and stability~\cite{Leung:2012vea}. Previous studies~\cite{Leung:2011zz,Leung:2012vea} showed that DM-admixed NSs can form equilibrium configurations distinct from those of standard NSs, potentially leaving observable imprints on macroscopic stellar properties~\cite{Ellis:2018bkr,Ellis:2017jgp}. In particular, Ref.~\cite{Ellis:2017jgp} emphasized that gravitational waves emitted during the merger of NSs with DM cores may help distinguish such systems from ordinary NS mergers. Related studies have also explored a variety of properties of hybrid admixed compact stars~\cite{Karkevandi:2021ygv,Kain:2021hpk,Deliyergiyev:2019vti,Leung:2022wcf,Das:2020ecp,Routaray:2022utr}. Recently, a fully relativistic treatment of polar (even‑parity) non‑radial oscillations for gravitationally coupled two‑fluid NSs,  with independently conserved currents and no direct microphysical coupling,  was developed in Ref.~\cite{Kumar:2026tgi}, focusing on the even‑parity sector where fluid motion couples directly to spacetime perturbations.  Additionally, Ref.~\cite{Dey:2025wcs} investigated NS $w$-modes in quarkyonic and DM-admixed equations of state using the polar perturbation formalism matched to the Zerilli equation. The complementary axial sector, which probes primarily the spacetime response of the equilibrium configuration, is addressed in this work.

The advent of gravitational-wave astronomy has turned compact-object mergers into a powerful probe of the dark sector. The detections of binary black-hole mergers by LIGO and Virgo~\cite{LIGOScientific:2016aoc}, as well as the multimessenger event GW170817 from a binary NS merger~\cite{LIGOScientific:2018cki}, have shown that gravitational-wave signals carry direct information about the internal structure and equation of state of compact objects. In systems containing a BS, or in BS–NS or BS–black hole mergers, the gravitational-wave signal may exhibit a distinctive late-time ringdown that reflects the nature of the underlying dark matter and its self-interactions. In this regime, the dominant signal is governed by QNMs, whose frequencies and damping times depend on the mass, spin, and composition of the remnant.  While the dominant ringdown emission is typically associated with the more strongly excited polar modes, the axial modes provide a complementary probe of the spacetime structure. The full QNM spectrum characterizes the linear gravitational response of the object, and any modification to the stellar interior should, in principle, leave imprints across the spectrum. Understanding the axial sector is therefore an important step toward a complete theoretical description of the ringdown spectrum of exotic compact objects.

Motivated by these considerations and by the well-studied BS model of Colpi \textit{et al.}~\cite{Colpi:1986ye}, we investigate the QNM spectrum of mixed NS-BS systems, which may form in dense DM environments, for example near DM spikes around supermassive black holes.  Our primary goal is to determine how the presence of self-interacting DM modifies the spacetime oscillation spectrum of compact stars. Using linear perturbation theory, we compute the axial QNM frequencies via the continued-fraction method~\cite{Leaver:1985ax,Leaver:1990zz} (see also Ref.~\cite{Kokkotas:1999bd}), study their dependence on the boson mass and self-coupling, and identify characteristic signatures associated with the DM component. These results complement recent studies of polar oscillations in two-fluid NSs and provide theoretical benchmarks for future studies of the oscillation spectra and gravitational-wave signatures of exotic compact objects.

This paper is organized as follows. In Sec.~\ref{sec2}, we present the background equations governing the equilibrium structure of the boson-fermion star configurations as well as the equations of state characterizing the baryonic matter and the DM. In Sec.~\ref{sec3}, we develop the formalism for axial perturbations and derive the corresponding perturbation equations. Section~\ref{sec4} is devoted to the numerical methods employed to solve the coupled system of equations and to derive the QNMs. In Sec.~\ref{sec5}, we discuss the QNM spectrum of boson-fermion stars and analyze the effects of the additional matter component on the oscillation properties. Finally, in Sec.~\ref{conclusion} we summarize our main results and present the concluding remarks. In this paper, we adopt natural units in which $G = c = 1$.

\section{Background equations}\label{sec2}
In this section, we derive the equations governing a spherically symmetric and static background, showing that the bosonic field can be recast as an effective perfect fluid.
\subsection{General relativistic formalism of two fluids}

The general structure of a compact object composed of two fluids is described by the Einstein equations,
\begin{eqnarray}\label{EEs}
    G_{\mu\nu}=8\pi T_{\mu\nu},
\end{eqnarray}
where the energy-momentum tensor is given by
\begin{equation}\label{TMUNU}
    T^{\mu\nu}=T^{\mu \nu}_{\rm NS}+T^{\mu \nu}_{\rm BS},
\end{equation}
with
\begin{eqnarray}
\quad  T^{\mu \nu}_{\rm NS}&=&  P_{\rm NS} g^{\mu\nu}+(P_{\rm NS}+\rho_{\rm NS})u^\mu_{\rm NS} u^\nu_{\rm NS},\\
  T^{\mu \nu}_{\rm BS}&=&  P_{\rm BS} g^{\mu\nu}+(P_{\rm BS}+\rho_{\rm BS})u^\mu_{\rm BS} u^\nu_{\rm BS},
\end{eqnarray}
where $u^\mu_{\rm NS}$ and $u^\mu_{\rm BS}$ denote the four-velocity vectors of the baryonic matter and bosonic matter, respectively, both satisfying the normalization conditions $g_{\mu\nu}u^\mu_{\rm NS}u^\nu_{\rm NS}=g_{\mu\nu}u^\mu_{\rm BS}u^\nu_{\rm BS}=-1$. For a two-fluid system, the total energy-momentum tensor is taken as the sum of the contributions from each fluid. In our case, we consider a bosonic fluid, which models DM, and a baryonic-matter fluid associated with the NS. We denote their energy densities by $\rho_{\rm BS}$ and $\rho_{\rm NS}$, respectively. We assume that the two components interact only through gravity, so that each fluid satisfies its own conservation law. Consequently, the covariant conservation of the total energy-momentum tensor,
\begin{equation}
\nabla_{\mu} T^{\mu \nu}=0,
\end{equation}
can be written as two independent equations,
\begin{equation}
\nabla_{\mu} T^{\mu \nu}_{\rm NS}=0,
\end{equation}
and
\begin{equation}
\nabla_{\mu} T^{\mu \nu}_{\rm BS}=0.
\end{equation}
This implies that each fluid separately conserves energy and momentum while evolving in the same spacetime geometry. Within this framework, we can model mixed compact configurations in which baryonic matter and DM coexist in hydrostatic equilibrium.
\subsection{Background equations for two admixed fluids}

We consider the static and spherically
symmetric spacetime with the line element,
\begin{equation}\label{eq:line_element}
    ds^2=-fdt^2+hdr^2+r^2(d\theta^2+\sin^2\theta d\phi^2),
\end{equation}
where  $f$ and $h$ are functions depending on the radial coordinate $r$. In this background, the energy density and the pressure of each fluid are considered to be functions of the radial coordinate $r$, i.e. $\rho_{\rm NS}\equiv\rho_{\rm NS}(r)$ and $\rho_{\rm BS}\equiv\rho_{\rm BS}(r)$.  From the field equations (\ref{EEs}), the background equations are written as follows
\begin{align}
    h^{-1}&=1-\frac{2m}{r},\\
    \frac{dm}{dr}&=4\pi r^2 \rho,\label{TOV1}\\
    \frac{f'}{2 f}&=\frac{m(r)+4\pi r^3 P}{r[r-2m(r)]},\label{TOV2}
    \end{align}
where $m(r)$ denotes the total enclosed mass at radial coordinate $r$. For an admixed star of baryonic matter and DM, we then have the total energy density $\rho$ and pressure $P$
\begin{align}
    \rho=\rho_{\rm NS}+\rho_{\rm BS},\quad P=P_{\rm NS}+P_{\rm BS}.
\end{align}
 The matter conservation of each fluid, in the metric \eqref{eq:line_element}, reads
    \begin{align}
    \frac{dP_{\rm NS}}{dr}&=-[P_{\rm NS}+\rho_{\rm NS}]\frac{f'}{2 f},\label{TOV3}\\
    \frac{dP_{\rm BS}}{dr}&=-[P_{\rm BS}+\rho_{\rm BS}]\frac{f'}{2 f}.\label{TOV4}
\end{align}
Physically, these equations describe two perfect fluids that coexist in the same spacetime and interact only through gravity. Each component is in hydrostatic equilibrium under the combined gravitational field generated by the total mass-energy content, but there is no direct momentum transfer between the baryonic and bosonic sectors. To solve for the structure, we only need to consider the system of ODEs for $m,P_{\rm NS},P_{\rm BS}$. Outside the star, both fluids vanish and the exterior spacetime reduces to Schwarzschild, which yields $m=M$ and $f=1-2M/r$, with $M$ a constant known as the ADM mass. From the above equations, one can verify that the total energy-momentum tensor is conserved
 \begin{eqnarray}
    \frac{dP}{dr}&=&-[P+\rho]\frac{f'}{2 f} 
 \end{eqnarray}
which corresponds to the matter conservation for the total fluid, showing that the radial variation of the total pressure is related to the spacetime curvature and the total energy density.
 
\subsubsection{Neutron matter equation of state}
We consider several fits of the neutron equation of state (EoS). Fortunately, they can all be expressed according to the following parametrization~\cite{Haensel:2004nu,Potekhin:2013qqa}:
\begin{align}\label{eq:NS_EOS}
    \zeta=&\frac{a_1+a_2\xi+a_3\xi^3}{1+a_4\xi}F[a_5(\xi-a_6)]+(a_7+a_8\xi)F[a_9(a_{10}-\xi)]\cr
    &+(a_{11}+a_{12}\xi)F[a_{13}(a_{14}-\xi)]+(a_{15}+a_{16}\xi)F[a_{17}(a_{18}-\xi)]\cr
    &+\frac{a_{19}}{1+[a_{20}(\xi-a_{21})]^2}+\frac{a_{22}}{1+[a_{23}(\xi-a_{24})]^2},
\end{align}
where we have defined the helpful function $F(x)\equiv \left(\mathrm{e}^x+1\right)^{-1}$, as well as the variables $\zeta\equiv\log(P/\mathrm{dyn}~\mathrm{cm}^{-2})$ and $\xi\equiv\log(\rho/\mathrm{g}~\mathrm{cm}^{-3})$. The values of the coefficients for each fit are provided in table~\ref{Tableai}.

The specific EoS models we adopt in this work are SLy, FPS, BSk20 and BSk21. SLy is based on a unified description of both the crust and the liquid core using the Skyrme-type effective nucleon-nucleon interaction SLy, which was specifically designed to describe very neutron-rich matter~\cite{Chabanat:1997qh,Douchin:2001sv}. On the other hand, FPS is another early unified EoS, derived from the energy-density functional of Pandharipande and Smith~\cite{Pandharipande:1976uol,Friedman:1981qw}.

The BSk20 and BSk21 EoSs belong to the Brussels--Montreal family of unified EoSs constructed from generalized Skyrme energy-density functionals fitted simultaneously to experimental nuclear masses and to microscopic pure-neutron-matter calculations~\cite{Goriely:2010bm,Potekhin:2013qqa}. These two models mainly differ in the stiffness of the symmetry energy and of the high-density neutron-matter sector: BSk20 is the softest and BSk21 is the stiffest~\cite{Potekhin:2013qqa,Chamel:2011aa}.

\subsubsection{Boson matter equation of state}

 We consider a model where DM consists of a complex scalar field $\phi$ with mass $m_\chi$ and a repulsive self-interaction $\lambda>0$. We closely follow the model of Colpi et al.~\cite{Colpi:1986ye}, where the Lagrangian density in curved spacetime is
\begin{equation}\label{Lphicom}
\mathcal{L}=\frac{1}{2}g^{\mu\nu}\nabla_\mu\phi^*\nabla_\nu\phi-\frac{1}{2}m_\chi^2|\phi|^2-\frac{\lambda}{4}|\phi|^4.
\end{equation}
We adopt the line element given in eq.~\eqref{eq:line_element} for spherically symmetric configurations and consider harmonic time dependence for the field, $\phi(r,t)=\Phi(r)e^{-i\omega_b t}$.

The structure of such bosonic configurations depends crucially on the strength of the self-interaction. Two distinct regimes can be identified, characterized by the dimensionless parameter
\begin{equation}
\Lambda \equiv \frac{\lambda}{4\pi}\frac{M_{\mathrm{Pl}}^2}{m_\chi^2},
\end{equation}
where $M_{\mathrm{Pl}}=1.2\times10^{19}$ GeV is the Planck mass. When $\Lambda \ll 1$, the self-interaction is negligible and the BS is supported primarily by gradient pressure, resembling a dilute state of condensed bosons. In the opposite limit, $\Lambda \gg 1$, the self-interaction dominates. From a particle physics perspective, natural values $\lambda\lesssim\mathcal{O}(1)$ and $m_\chi\ll M_{\mathrm{Pl}}$ place us precisely in the strong-coupling regime $\Lambda\gg 1$. In this limit, first studied in Ref.~\cite{Colpi:1986ye} and subsequently employed in later investigations of strongly self-interacting BSs~\cite{Gleiser:1988rq,Ryan:1996nk}, the quartic self-interaction dominates over the scalar-field gradient energy throughout most of the stellar interior. Consequently, the Einstein-Klein-Gordon system admits a hydrodynamic description in which the scalar condensate behaves as an effective perfect fluid obeying an algebraic equation of state. Rather than solving the full Einstein-Klein-Gordon equations and subsequently taking the strong-coupling limit, we adopt this effective fluid description from the outset, following Ref.~\cite{Kamenetskaia:2022lbf}. This formulation is equivalent to the classical self-interacting BS model of Ref.~\cite{Colpi:1986ye} within the regime $\Lambda\gg1$, while being particularly convenient for constructing two-fluid NS/DM configurations. Within this hydrodynamic formulation, the equation of state follows from the Einstein-Klein-Gordon system in the strong-coupling limit. To this end, we introduce the rescaled variables
\begin{equation}\label{eq:rescaled_variables}
x_* \equiv m_\chi r\Lambda^{-1/2}, \quad \sigma_* \equiv \sqrt{4\pi}\frac{\Phi}{M_{\mathrm{Pl}}}\Lambda^{1/2}, \quad f_* \equiv \frac{f}{\Omega^2}, \quad \Omega \equiv \frac{\omega_b}{m_\chi},
\end{equation}
and following the same procedure as in~\cite{Kamenetskaia:2022lbf}. One then finds that the field amplitude becomes algebraically related to the metric function:
\begin{equation}
\sigma_* = \left(\frac{1}{f_*}-1\right)^{1/2}. \label{eq:sigmaBrelation}
\end{equation}
 Equation~(\ref{eq:sigmaBrelation}) is the characteristic algebraic relation that emerges in the hydrodynamic limit, allowing the energy density and pressure to be written solely in terms of the metric function $f_*$ as 
\begin{align}
		&\rho_{*}=\frac{1}{16\pi}\left(\frac{3}{f_{*}}+1\right)\left(\frac{1}{f_{*}}-1\right), \cr
		&P_{*}=\frac{1}{16\pi}\left(\frac{1}{f_{*}}-1\right)^2,\cr
		&f_{*}=\frac{f}{\Omega^2}.
\end{align}
Eliminating $f_*$ between these expressions yields the equation of state for the bosonic matter in the strong-coupling regime:
\begin{equation}
P_* = \frac{1}{36\pi}\left(\sqrt{1+12\pi\rho_*}-1\right)^2. \label{eq:BosonEoS}
\end{equation}
This relation has two asymptotic limits. At low densities ($\rho_*\ll 1$), it expands to $P_* \approx \pi\rho_*^2$, a polytropic equation of state characteristic of a repulsive self-interaction. At high densities ($\rho_*\gg 1$), it goes to $P_* \propto \rho_*$, corresponding to a stiff fluid.  Within the hydrodynamic description, the Einstein-Klein-Gordon system becomes mathematically equivalent to the Tolman-Oppenheimer-Volkoff (TOV) equations for a perfect fluid obeying Eq.~(\ref{eq:BosonEoS}) (see Refs.~\cite{Colpi:1986ye,Kamenetskaia:2022lbf} for the derivation). The stellar structure is therefore governed by
\begin{subequations}\label{eq:DSStellar_Structure}
		\begin{align}
			& \frac{dm_{*}}{dx_{*}}=4\pi x_{*}^2\rho_{*}, \label{DSStellarStructureA} \\
			&\frac{dP_{*}}{dx_{*}}=-\frac{1}{x_{*}^2}\left(4\pi P_{*}x_{*}^3+m_{*}\right)\left(\rho_{*}+P_{*}\right)\left(1-\frac{2m_{*}}{x_{*}}\right)^{-1},\label{DSStellarStructureB}\\
            &h(x_*)\equiv\left(1-\frac{2m_*(x_*)}{x_*}\right)^{-1},
            \label{DSStellarStructureC}
		\end{align}
	\end{subequations}
where eq.~\eqref{DSStellarStructureA} is the mass equation and eq.~\eqref{DSStellarStructureB} is the TOV equation. In order to couple this DM model with the neutron matter, it is convenient to undo the various changes of variables in the derivation of eq.~\eqref{eq:BosonEoS}, and express the dimensionful enclosed mass, radius, density and pressure ($m$, $r$, $\rho$ and $p$) in terms of their dimensionless counterparts. We obtain
\begin{subequations} \label{eq:Unit_Conversion}
		\begin{align} 
			& m=0.46M_\odot\lambda^{\frac{1}{2}}\left(\frac{m_\chi}{1~\rm GeV}\right)^{-2}m_*, \label{Mass Conversion2} \\
			& r=0.68~{\rm km}~\lambda^{\frac{1}{2}}\left(\frac{m_\chi}{1~\rm GeV}\right)^{-2}x_*, \label{Length Conversion2} \\
			&\rho=2.9\times10^{18}\ \frac{\mathrm{gr}}{\mathrm{cm}^3}~\lambda^{-1}\left(\frac{1\ \mathrm{GeV}}{m_\chi}\right)^{-4}\rho_{*} ,\label{Density Conversion} \\
			&P=2.6\times10^{38}\ \mathrm{Pa}~\lambda^{-1}\left(\frac{1\ \mathrm{GeV}}{m_\chi}\right)^{-4}P_{*}, \label{Pressure Conversion} 
		\end{align}
	\end{subequations}
leading to the dimensionful boson EoS
\begin{equation}\label{eq:EoSBS}
    P(\rho)=\frac{\rho_0 c^2}{36\pi}\left(\sqrt{1+12\pi\left(\frac{\rho}{\rho_0}\right)}-1\right)^2,
\end{equation}
where we define $\rho_0\equiv2.92\times10^{18}\ \frac{\mathrm{g}}{\mathrm{cm}^3}~\lambda^{-1}\left(\frac{m_\chi}{1\ \mathrm{GeV}}\right)^{4}$. Finally, we note from eq.~\eqref{eq:Unit_Conversion} that BS solutions are degenerate in the microphysical parameters. Namely, any combination of $\lambda$ and $m_\chi$, such that the product $\lambda\, m_\chi^{-4}$ is the same will yield the same structure. With this in mind, we will henceforth fix the self-coupling to $\lambda=0.01$ and only vary the boson mass to consider different configurations. Furthermore, we will consider BS solutions that are similar in mass to NS solutions. To this end, we pick the benchmark values of $m_\chi=0.04,0.08,0.10,0.12~\rm GeV$, which correspond, respectively, to maximal BS mass of $M_{\rm BS}=6.25,1.56,1.00,0.69~M_\odot$.

\section{Axial perturbations}\label{sec3}
In this section, we present the axial perturbation equations, which allow us to determine the QNMs. Axial perturbations are particularly useful because, for a non-rotating perfect-fluid star, they do not couple directly to pressure and density perturbations in the same way as polar modes. As a result, the axial spectrum probes primarily the spacetime curvature generated by the equilibrium matter distribution.

In order to study axial perturbations, we consider a small perturbation $h_{\mu\nu}$ around a static and spherically symmetric metric $g^{0}_{\mu\nu}$ as 
\begin{equation}
    g_{\mu\nu}=g^{0}_{\mu\nu}+h_{\mu\nu}.
\end{equation}
The perturbation $h_{\mu\nu}$  can be decomposed  into two sectors with  different parity, which are  decoupled at the linear level.  In the odd-parity sector, the perturbations $h_{\mu\nu}$  are decomposed  into  spherical harmonics $Y_{lm}(\theta,\phi)$ as
\begin{eqnarray}
    h_{at}&=& r^2\sum_{lm} H_{0}^{lm}(r,t)\sin\theta\,\epsilon_{ab}\nabla^b Y_{lm}\\
    h_{ar}&=&r^2\sum_{lm} H_{1}^{lm}(r,t)\sin\theta\,\epsilon_{ab}\nabla^b Y_{lm},\\
 h_{ab}&=&\frac{1}{2}r^2 \sum_{lm} Q^{lm}(r,t)\sin\theta(\epsilon_{a}^{c}\nabla_c\nabla_b Y_{lm}+\epsilon_{b}^{c}\nabla_c\nabla_a Y_{lm}),
\end{eqnarray}
where $H_0$, $H_1$ and $Q$ are functions of the radial and the temporal components.  The indices  $\{a,b\}\equiv\{\theta,\phi\}$  are raised or lowered using the unit 2-sphere metric and $\epsilon_{ab}$ is the antisymmetric tensor with $\epsilon_{\theta\phi}=-1$.

Under an infinitesimal gauge transformation $x^\mu \to x^\mu + \xi^\mu$, the axial-sector gauge vector is chosen such that $\xi^t = \xi^r = 0$, while its angular components take the form
\begin{eqnarray}
\xi^a = r^2 \sum_{lm} \Lambda^{lm}(r,t)\,\sin\theta\, \epsilon^{a}{}_{b}\nabla^b Y_{lm}.
\end{eqnarray}
This transformation changes the perturbation amplitudes according to
\begin{eqnarray}
    H_{0}^{lm}  \rightarrow  H_{0}^{lm}+\dot{\Lambda}^{lm},\quad
    H_{1}^{lm} \rightarrow  H_{1}^{lm}+\Lambda^{lm}\, ',\quad
    Q^{lm}  \rightarrow  Q^{lm}+2\Lambda^{lm}\,.
\end{eqnarray}
We fix the gauge by choosing $\Lambda^{lm}=-Q^{lm}/2$, which sets $Q^{lm}=0$. For axial perturbations, the perturbed four-velocity  of the fluid is of the form 
\begin{eqnarray}
\label{4-velocity}
u^\mu_{\rm NS}=\bar{u}^\mu_{\rm NS}+\delta u^\mu_{\rm NS}\,, \qquad u^\mu_{\rm BS}=\bar{u}^\mu_{\rm BS}+\delta u^\mu_{\rm BS}\,, \qquad \bar{u}^\mu_{\rm NS}=\bar{u}^\mu_{\rm BS}=\frac{\delta^\mu_0}{f^{1/2}}\,.
\end{eqnarray} 
The nonzero components can be decomposed as 
\begin{eqnarray}
\delta u^{a}_{\rm NS}=r^2 \sum_{lm} v^{lm}_{\rm NS}(r,t)\sin\theta\,\epsilon^{a}_{\  b}\nabla^b Y_{lm},\\
  \delta u^{a}_{\rm BS}=r^2 \sum_{lm} v^{lm}_{\rm BS}(r,t)\sin\theta\,\epsilon^{a}_{\  b}\nabla^b Y_{lm}.
\end{eqnarray}
In the following, without loss of generality, we omit the indices $l$ and $m$ from the metric  functions and the velocity in order to lighten the notation.
Choosing the Regge-Wheeler gauge \cite{Regge:1957td}, the perturbed metric is written as follows
\begin{eqnarray}
\label{perturbed_metric}
ds^2&=& -f \,dt^2+h\,dr^2+r^2\left(d\theta^2 + \sin^2 \theta d\phi^2\right)\nonumber\\
& &+2 r^2\sin \theta \partial_\theta Y_{l0}(\theta)\left(H_0(r,t)dtd\phi +H_1(r,t)drd\phi \right),
\end{eqnarray}
where both  perturbations $H_0(r,t)$ and $H_1(r,t)$ depend implicitly on $l$.

If we substitute the perturbed metric \eqref{perturbed_metric} and the perturbed four-velocity \eqref{4-velocity} into the energy-momentum tensor \eqref{TMUNU}, we find that the non-vanishing perturbed components of the tensor,  at linear order in the perturbations, are given by
\begin{eqnarray}
\label{T_tphi}
\qquad 
\delta u^\mu_{\rm x} &=& \left\lbrace 0,0,0,v_{\rm x}\frac{\partial_\theta Y_{l0}(\theta)}{f^{\frac{1}{2}}\sin\theta  } \right\rbrace\,\\
\delta T^{t\phi}_{\rm x}&=&\frac{\partial_\theta Y_{l0}(\theta)}{\sin \theta\, f}\left[ P_{\rm x} H_0+(\rho_{\rm x}+P_{\rm x})v_{\rm x}\right],\\
\label{T_ttheta}
\delta T^{r\phi}_{\rm x}&=&-\frac{\partial_\theta Y_{l0}(\theta)}{\sin \theta\, h}P_{\rm x} H_1.
\end{eqnarray}
where the subscript ${\rm x}$ stands for either ${\rm NS}$ or ${\rm BS}$. Then, using the perturbed matter conservation equation, one can find that 
\begin{eqnarray}
\label{H0}
v_{\rm BS}=v_{\rm NS}=-H_0\,.
\end{eqnarray}
As for a single fluid, this result indicates that no independent degree of freedom is associated with the fluid in the axial sector. Substituting the expressions of $\delta T^{t\phi}_{\rm x}$ and $\delta T^{r\phi}_{\rm x}$ into the field equations~(\ref{EEs}), and making use of eq.~(\ref{H0}), we obtain two independent partial differential equations. By integrating over $\theta$, the angular dependence is removed. The perturbation equations in the odd-parity sector, expressed in Fourier space ($H(r,t)\to\int dt\, e^{i\omega t} H(r)$), take the form
 \begin{eqnarray}
     H_0'&=& \frac{i \, H_1 }{r^2 \omega }\left( f(2-\ell(\ell+1))+r^2 \omega ^2\right)\\
     H_1'&=& \frac{1}{2} \left(H_1 \left(\frac{h'}{h}-\frac{f'}{f}-\frac{4}{r}\right)+H_0\frac{2 i \omega\,  h }{f}\right)
 \end{eqnarray}
 From the above equations  and by defining $\Psi=r H_1/h$, one can obtain the Regge–Wheeler equation \cite{Chandrasekhar:1991fi} as follows
\begin{equation}\label{eq:axial_perturbation}
    \frac{d^2\Psi}{dr_*^2}+ \left[\omega ^2-V(r)\right]\Psi =0,
\end{equation}
where $r_*=\int\sqrt{h/f}~dr$ is the tortoise coordinate and the potential reads
\begin{equation*}
    V(r)=f\left[\frac{\ell (\ell+1)}{r^2}+4\pi(\rho-P)-\frac{6m}{r^3}\right].
\end{equation*}
The effective axial potential has the same Regge–Wheeler structure of General Relativity; however, the matter source consists of two independent components. The bosonic contribution modifies the potential through the density, pressure and mass terms, which depend purely on the scalar-field self-interaction potential. Consequently, deviations from the NS spectrum arise from the modified internal matter composition.

The solution of eq.~\eqref{eq:axial_perturbation} with the boundary conditions of regularity at the center and purely outgoing radiation at infinity (see Sec.~\ref{sec:axial_numerical} for more details) yields a discrete spectrum of complex eigenfrequencies $\omega = \mathrm{Re}[\omega] + i\,\mathrm{Im}[\omega]$. The real part gives the angular oscillation frequency, while the imaginary part (or, equivalently, the damping time $\tau \equiv 1/|\mathrm{Im}[\omega]|$) determines the gravitational wave decay. These complex eigenfrequencies are the axial QNMs of the star. Equation~\eqref{H0} shows that the fluid perturbation $v$ is algebraically tied to the metric perturbation $H_0$, which means it does not represent an independent degree of freedom. As a result, the fluid is simply dragged along by the spacetime oscillation, and the axial QNMs are pure gravitational modes. These are the so-called $w$-modes~\cite{Chandrasekhar:1991fi,Kokkotas:1999bd} (sometimes called ``wave modes'' or ``spacetime modes'') to distinguish them from the fluid-dominated $f$-, $p$-, and $g$-modes that appear in the polar sector.

The axial $w$-mode spectrum splits into different families. For every stable stellar configuration there exists a fundamental branch, referred to as the $w$I branch~\cite{Blazquez-Salcedo:2012hdg}. Within this branch the modes are ordered by the magnitude of the damping time: the fundamental mode ($w$I$_0$) has the lowest oscillation frequency and the longest damping time (smallest magnitude of $\mathrm{Im}[\omega]$), while the first overtone ($w$I$_1$) has a higher frequency and a shorter damping time, and so on. All wI modes are present independently of the compactness of the star, making them universal signatures of compact object oscillations. In the remainder of this work we will restrict our analysis to the quadrupolar ($l=2$) $w$I modes, with particular attention given to the fundamental, as it is the mode most likely to be excited in astrophysical events and to fall within the sensitivity window of current ground-based gravitational wave detectors.

\section{Numerical solution}\label{sec4}
In this section, we present the numerical analysis of the mixed two-fluid system composed of baryonic matter and bosonic DM within the framework of general relativity. By solving the equilibrium equations for different central densities and model parameters, we investigate the impact of the DM component on the global properties of the star, including the MR relation, density profiles and QNMs.  For numerical convenience, we introduce a set of dimensionless variables defined as follows:
\begin{eqnarray}
 r= r_0 \hat{r}\,,\quad
    \rho=\rho_{\rm typ}\hat{\rho}\,,\quad
    P=\rho_{\rm typ} \hat{P}\,,
\end{eqnarray}
with
\begin{equation}
 n_0=0.1 \,{\rm fm^{-3}}\,, \qquad  \rho_{\rm typ}=m_b\, n_0=1.6749\times 10^{14} {\rm \,g/cm^3}\,, \qquad   r_0=\frac{1}{\sqrt{\rho_{\rm typ}}}=89.667\;{\rm km}\,.
\end{equation}
where $m_b$ is the nucleon mass at rest.
\subsection{Background solutions}
The stellar configurations are obtained by solving  equations~\eqref{TOV1}-\eqref{TOV4} numerically, starting from the center of the star to large value of $r$. At the center of the star, we develop a Taylor expansion around $r=0$ and then we solve Eqs.~(\ref{TOV1}), (\ref{TOV2}), (\ref{TOV3}) and (\ref{TOV4}). For a mixed system, the Taylor expansions at the center are
 \begin{eqnarray}
     f/f_c&=& 1+\frac{4\pi}{3}\left[3(P_{\rm NS,c}+P_{\rm BS,c})+(\rho_{\rm NS,c}+\rho_{\rm BS,c})\right]\,r^2,\\
     m&=&\frac{4\pi}{3}(\rho_{\rm NS,c}+\rho_{\rm BS,c})r^3,\\
     P_{\rm NS}&=& P_{\rm NS,c}-\frac{2\pi}{3}\left[3(P_{\rm NS,c}+P_{\rm BS,c})+(\rho_{\rm NS,c}+\rho_{\rm BS,c})\right]\left(P_{\rm NS,c}+\rho_{\rm NS,c}\right)\,r^2,\\
     P_{\rm BS}&=& P_{\rm BS,c}-\frac{2\pi}{3}\left[3(P_{\rm BS,c}+P_{\rm NS,c})+(\rho_{\rm BS,c}+\rho_{\rm NS,c})\right]\left(P_{\rm BS,c}+\rho_{\rm BS,c}\right)\,r^2,
 \end{eqnarray}
where $f_c$, $\rho_{\rm NS,c}$, and $\rho_{\rm BS,c}$ denote the central values of $f$, $\rho_{\rm NS}$, and $\rho_{\rm BS}$, respectively. The central pressures $P_{\rm NS,c}$ and $P_{\rm BS,c}$ are determined from their respective equations of state, while $f_c$ is chosen to satisfy the condition $f(\infty)=1$.

\begin{figure}[h!]
	\centering
	\includegraphics[width=.48\textwidth]{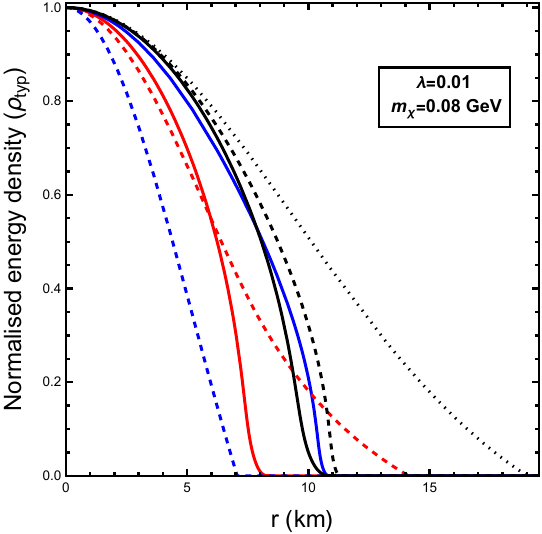}
    \includegraphics[width=.48\textwidth]{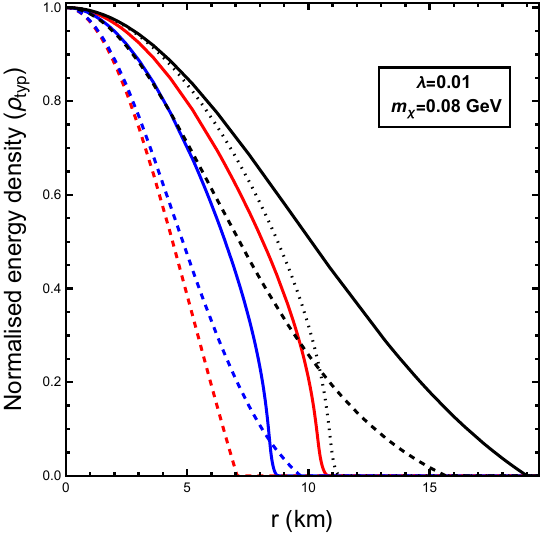}
	\caption{Radial profiles of the neutron-matter density $\rho_{\mathrm{NS}}$ and bosonic DM density $\rho_{\mathrm{BS}}$ for configurations obtained with the SLy equation of state. \textit{Left panel:} Density profiles normalized to their central values, $\rho_{\mathrm{NS}}/\rho_{\mathrm{NS},c}$ (solid lines) and $\rho_{\mathrm{BS}}/\rho_{\mathrm{BS},c}$ (dashed lines), for fixed bosonic central density $\rho_{\mathrm{BS},c}=2\rho_{\mathrm{typ}}$. The neutron central density $\rho_{\mathrm{NS},c}$ is varied, with red and blue corresponding to $\rho_{\mathrm{NS},c}=3\rho_{\mathrm{typ}}$ and $\rho_{\mathrm{NS},c}=6\rho_{\mathrm{typ}}$, respectively. The black dotted line shows the pure BS configuration, while the black solid and dashed lines show the corresponding pure NS configurations for $\rho_{\mathrm{NS},c}=3\rho_{\mathrm{typ}}$ and $\rho_{\mathrm{NS},c}=6\rho_{\mathrm{typ}}$, respectively. \textit{Right panel:} Density profiles $\rho_{\mathrm{NS}}/\rho_{\mathrm{NS},c}$ (solid lines) and $\rho_{\mathrm{BS}}/\rho_{\mathrm{BS},c}$ (dashed lines) for fixed neutron central density $\rho_{\mathrm{NS},c}=6\rho_{\mathrm{typ}}$. The bosonic central density is varied, with red and blue corresponding to $\rho_{\mathrm{BS},c}=2\rho_{\mathrm{typ}}$ and $\rho_{\mathrm{BS},c}=6\rho_{\mathrm{typ}}$, respectively. The black dotted line shows the pure NS configuration, while the black solid and dashed lines show the corresponding pure BS configurations for $\rho_{\mathrm{BS},c}=2\rho_{\mathrm{typ}}$ and $\rho_{\mathrm{BS},c}=6\rho_{\mathrm{typ}}$, respectively.}
	\label{rrho}
\end{figure}

The integration  performed from $r=0$ outwards leads to the definition of two distinct radii associated with each fluid. The neutron radius $R_{\rm NS}$ is defined by the vanishing of the baryonic pressure ($P_{\mathrm{NS}}(R_{\mathrm{NS}})=0$) and corresponds to the boundary of the baryonic component. On the other hand, the bosonic radius $R_{\rm BS}$ is defined by the vanishing of the bosonic pressure ($P_{\mathrm{BS}}(R_{\mathrm{BS}})=0$)~\cite{Leung:2011zz,Xiang:2013xwa}. Depending on the relative central densities, either component may extend farther outward. We note that in some cases the vanishing pressure of bosonic matter can occur before the baryonic pressure vanishes. In these cases, we keep integrating the TOV equations  to  the radius corresponding to the ordinary matter radius $R_{\mathrm{NS}}$. The total gravitational mass $M$ is obtained from the  value of the mass function at large radius, $M=m(\mathrm{max}[R_{\rm NS},R_{\rm BS}])$. The  presence of two radii provides a natural description of mixed two fluids.

The radial profiles shown in the left panel of fig.~\ref{rrho} illustrate the impact of varying the neutron-matter content while keeping the bosonic matter central density fixed. For $\rho_{\mathrm{NS},c}=3\rho_{\mathrm{typ}}$, the baryonic matter dominates the inner region of the star, while the bosonic density extends to larger radii. On the other hand, when the neutron central density is increased to $\rho_{\mathrm{NS},c}=6\rho_{\mathrm{typ}}$, the largest size is that of the baryonic component. In addition, the neutron density profile is modified by the additional gravitational contribution of the bosonic fluid, as can be seen by comparing the mixed configurations with the pure NS case (black solid and dashed lines).

The left panel also shows that the bosonic component becomes progressively more compact as the relative neutron content increases. In the absence of neutron matter, the pure BS configuration is extended, with a radius of $\sim19\,\mathrm{km}$ for the parameter choice considered here. As $\rho_{\mathrm{NS},c}/\rho_{\mathrm{BS},c}$ increases, the bosonic radius decreases, and for sufficiently large neutron central density it can even become smaller than the NS radius. This behaviour implies that the bosonic contribution to the total mass also decreases, since the dark component occupies a smaller volume and therefore contributes less to the overall energy budget of the star.

Additionally, the presence of the bosonic component compresses the neutron distribution relative to the corresponding pure NS configurations. This is evident from the comparison between the black solid and red solid curves, as well as between the black dashed and blue solid curves. In both cases, the neutron-matter radius is reduced once the bosonic fluid is included, showing that the dark component modifies the baryonic structure through its gravitational backreaction. The magnitude of this reduction depends on the relative central densities of the two fluids. When $\rho_{\mathrm{NS},c}/\rho_{\mathrm{BS},c}$ is larger, the baryonic component dominates the configuration and the change in the neutron radius is less pronounced, as seen in the comparison between the black dashed and blue solid curves. By contrast, when the bosonic fraction is relatively larger, the contraction of the neutron-matter distribution becomes more significant.

This behavior is also confirmed in the right panel, where we show the effect of varying the neutron matter for fixed values of the neutron central density. As the fraction $\rho_{\mathrm{NS},c}/\rho_{\mathrm{BS},c}$ decreases, the BS density profile approaches the pure BS case and the NS component becomes smaller and smaller. In fact, for $\rho_{\mathrm{BS},c}=2\rho_{\mathrm{typ}}$, we observe $R_{\rm NS}>R_{\rm BS}$, while we see that $R_{\rm NS}<R_{\rm BS}$ when $\rho_{\mathrm{BS},c}=6\rho_{\mathrm{typ}}$. The comparison with the pure NS and pure BS configurations shows that the inclusion of an additional matter component significantly modifies the radial extension of the star. Nevertheless, the qualitative shape of each density profile remains largely unchanged. In particular, the bosonic density profile preserves its characteristic convex behavior near the stellar surface, whereas the neutron-matter profile remains concave. This indicates that the presence of the second fluid primarily affects the global gravitational potential and the spatial extent of each component, rather than the intrinsic structure of their profiles. The origin of this behavior can be traced to the EoSs governing each fluid. Within the two-fluid framework considered here, the neutron and bosonic sectors interact only through gravity, so the EoS of each component is not modified by the presence of the other fluid. Consequently, the characteristic profile shape associated with each EoS is preserved, even though the overall equilibrium configuration is substantially altered.

\begin{figure}[h!]
	\centering
	\includegraphics[width=.48\textwidth]{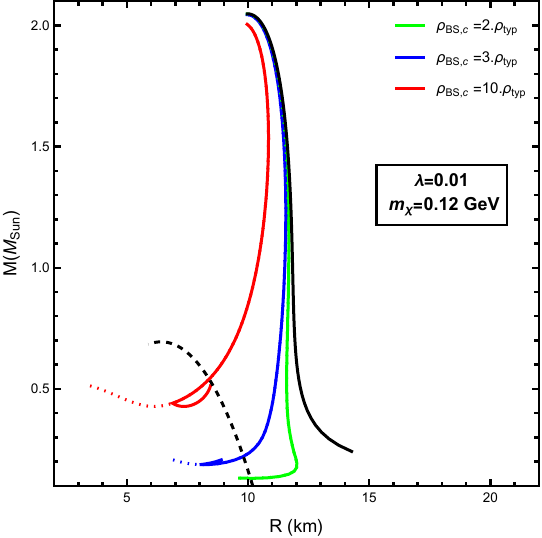}
    \includegraphics[width=.48\textwidth]{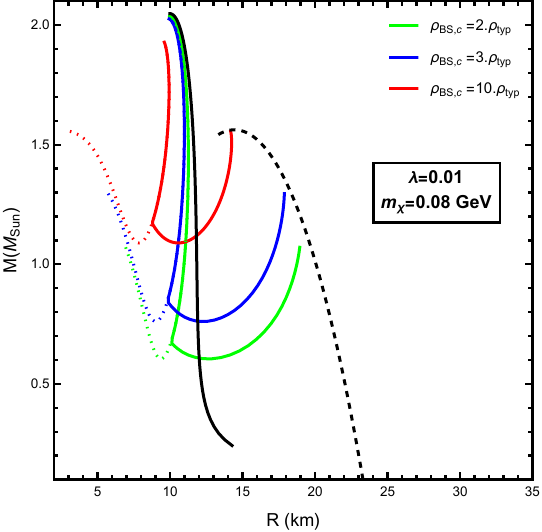}
	\caption{MR relations for boson–fermion stars with the SLy equation of state and two representative boson masses, $m_\chi=0.12~\rm GeV$ (left) and $m_\chi=0.08~\rm GeV$ (right). The solid black curve is the pure NS sequence, while the dashed black curve is the pure BS sequence. Colored solid curves are sequences with a fixed central boson density, as indicated in each panel, while the neutron central density $\rho_{\rm NS,c}$ is varied from $1.5\rho_{\rm typ}$ up to the maximum mass. The dotted colored curves show the same sequences but with the mass plotted against the neutron‑matter radius $R_{\rm NS}$ instead of the total radius $R=\mathrm{max}(R_{\rm NS},R_{\rm BS})$. When the neutron component is larger ($R_{\rm NS}>R_{\rm BS}$), the solid and dotted curves coincide; when the bosonic component dominates ($R_{\rm BS}>R_{\rm NS}$), the solid curve follows the boson radius while the dotted curve shows the (smaller) neutron radius.}
	\label{MR}
\end{figure}
It is well known that mass and radius are the two most important observable properties of a compact star. For a single-fluid star, the solutions of the TOV equations form a one-parameter family: varying the central density traces out a single MR curve. In a two-fluid system, however, the parameter space is larger because the central densities of the neutron and bosonic components can be chosen independently. In this work we construct MR relations by fixing the central energy density of the bosonic matter, $\rho_{\rm BS,c}$, and varying the central density of the ordinary nuclear matter, $\rho_{\rm NS,c}$. The stellar radius $R$ is taken to be the largest of the two component radii, that is,
\begin{eqnarray}
    R=\mathrm{max}[R_{\rm NS},R_{\rm BS}].
\end{eqnarray}
It is convenient to use  $\rho_{\rm NS,c}$ and  $\rho_{\rm BS,c}$  as primary parameters because the central energy densities constitute natural parameters for solving the TOV equations, for a given equation of state and bosonic field configuration. By contrast, global quantities such as the dark matter fraction~\cite{DiGiovanni:2021ejn} are obtained only after computing the background solutions and therefore are less suitable as primary parameters.

Such alternative parametrizations may be more appropriate for modeling specific astrophysical formation scenarios. However, in our case, rather than focusing on a particular formation mechanism, we aim to characterize the axial QNM spectrum via the complete family of equilibrium solutions admitted by the relativistic two-fluid equations. The choice of fixed central bosonic density therefore provides a convenient parametrization of this solution space, allowing us to follow continuously the transition from NS-dominated to BS-dominated configurations.

To illustrate the impact of the two central densities, we present in Fig.~\ref{MR} the MR relations for the two-fluid system, using two representative boson masses $m_\chi$. Each colored curve is obtained by fixing the bosonic central density, $\rho_{\mathrm{BS},c}$, and varying the neutron central density, $\rho_{\mathrm{NS},c}$, from a minimum value of $1.5\rho_{\rm typ}$ up to the point where the total mass reaches its maximum for that fixed $\rho_{\mathrm{BS},c}$. The lower bound $1.5\rho_{\rm typ}$ is a practical cutoff, namely for smaller values, the baryonic component becomes extremely dilute and the configuration is essentially a pure BS. Extending the curves all the way to $\rho_{\mathrm{NS},c}=0$ would produce a very long tail towards small NS radius, that would not add relevant information. We point out that each sequence converges to the corresponding pure BS solution (dashed black curve) in this limit, but we truncate the plots at $1.5\rho_{\rm typ}$ for clarity. Increasing $\rho_{\mathrm{BS},c}$ moves the entire sequence closer to the pure BS branch, as expected.

The solid colored lines show the gravitational mass $M$ versus the total stellar radius $R = \max(R_{\rm NS},R_{\rm BS})$, where $R_{\rm NS}$ and $R_{\rm BS}$ are the radii at which the respective pressures vanish. The dotted lines of the same color show the same mass but plotted against the neutron-matter radius $R_{\rm NS}$ alone. When the neutron component extends further than the bosonic one ($R_{\rm NS}>R_{\rm BS}$), the total radius equals $R_{\rm NS}$, hence the solid and dotted curves overlap. When the bosonic component is larger ($R_{\rm BS}>R_{\rm NS}$), the solid curve follows the larger boson radius while the dotted curve continues to track the shrinking neutron radius.

An interesting feature is that for a fixed $\rho_{\mathrm{BS},c}$, the solid MR curves can present a non-monotonic behavior, so that the same total mass may correspond to two different total radii. This is an effect that arises from the two-fluid description we consider here, and is due to the competition between the two components as the neutron density increases. At low $\rho_{\mathrm{NS},c}$ the star is BS-dominated (large $R_{\rm BS}$), whereas at high $\rho_{\mathrm{NS},c}$ it becomes NS-dominated (large $R_{\rm NS}$). Near the transition the total radius may first decrease and then grow, producing the loop present in this figure. The sequences are then understood as the continuous interpolation between the pure NS and pure BS limits, with the relative contribution of each fluid determining the global properties. Additionally, all configurations shown lie on the stable branch of the equilibrium sequence, i.e. they satisfy $dM/d\rho_c>0$ for fixed $\rho_{\mathrm{BS},c}$. A full stability analysis would require studying radial perturbations, which is beyond the scope of this work; however, our focus on the stable branch ensures that the backgrounds used for the QNM computation are dynamically well-behaved.

In Fig.~\ref{MR}, we show results for two values of $m_\chi$ in separate panels, where we observe a clear distinction in the MR relation. For $m_\chi=0.12\,\mathrm{GeV}$, the configurations have approximately the same compactness as a NS, with radii not exceeding the NS radius. We also observe that the boson-fermion stars stay closer to the NS branch for high values of the neutron central density. However, for $m_\chi=0.08\,\mathrm{GeV}$, the radius is much larger, reaching up to $\sim 20$-$25$ km. This difference is a direct consequence of the difference between the BS and NS radii, implying that a lighter boson mass generates more extended structures, as expected from eq.~\eqref{Length Conversion2}.

Finally, we point out that the equilibrium sequences shown in Fig.~\ref{MR} interpolate continuously between two physically distinct limits. At one end, the bosonic component is subdominant and the configurations correspond to DM‑admixed NSs, which have been widely studied as a consequence of DM capture or accretion onto ordinary NSs. At the opposite end, the baryonic component becomes subdominant and the solutions are more naturally interpreted as BSs containing a smaller component of baryonic matter. Such configurations are also of physical interest, for example if bosonic compact objects form prior to baryonic accretion or if primordial bosonic compact objects subsequently capture ordinary matter. The choice to explore the full range of admixed configurations, including the BS‑dominated limit, is also motivated from a theoretical standpoint, as the two‑fluid system admits a continuous family of equilibrium solutions that interpolate between the pure NS and pure BS limits. Understanding both the equilibrium solutions and the QNM spectrum (see the next section) across this entire parameter space is essential for characterizing the gravitational response of such objects.

\subsection{Axial perturbation solutions}\label{sec:axial_numerical}
Having constructed the equilibrium two-fluid configurations and analyzed their macroscopic properties, we now turn to the computation of the axial QNMs and the numerical procedure used to determine their complex eigenfrequencies.
\subsubsection{Inner integral}
For numerical integration, it is more convenient to rewrite the perturbation equation~\eqref{eq:axial_perturbation} in terms of the radial coordinate instead of the tortoise coordinate. We define the helpful function $s(r)\equiv\sqrt{h/f}$, which allows us to write
\begin{equation}\label{eq:axial_inner}
    \frac{d^2\Psi}{dr^2}-\frac{ds/dr}{s}\frac{d\Psi}{dr}+s^2[\omega^2-V(r)]\Psi=0.
\end{equation}
Performing a Frobenius series and imposing regularity at the center, we get the boundary condition for $\Psi$ as
\begin{equation*}
    \Psi(r\to0)=a_{\rm in}r^{\ell+1}.
\end{equation*}
We are free to choose $a_{\rm in}$, for numerical convenience we choose to define as $a_{\rm in}=i(\omega/4)^{\ell-1}$. We then have the boundary conditions at the center:
\begin{equation*}
    \Psi(r\to0)=i\left(\frac{\omega}{4}\right)^{\ell-1}r^{\ell+1},\ \left(\frac{d\Psi}{dr}\right)_{r\to0}=i\left(\frac{\omega}{4}\right)^{\ell-1}(\ell+1)r^{\ell}.
\end{equation*}

\subsubsection{Outer integral}
Here, we make use of the continued-fraction method~\cite{Leaver:1985ax}. For this, it is advantageous to rewrite the axial perturbation equation~\eqref{eq:axial_perturbation} outside the star in terms of the radial coordinate to obtain
\begin{equation*}
    \left(1-\frac{2M}{r}\right)^2\frac{d^2\Psi}{dr^2}+\left(1-\frac{2M}{r}\right)\frac{2M}{r^2}\frac{d\Psi}{dr}+[\omega^2-V(r)]\Psi=0,
\end{equation*}
where the potential is simplified into the form
\begin{equation}
    V(r)=\left(1-\frac{2M}{r}\right)\left[\frac{\ell(\ell+1)}{r^2}-\frac{6M}{r^3}\right].
\end{equation}
We now propose the ansatz of the homogeneous Regge-Wheeler equation in a power-series expansion of the form
\begin{equation}\label{eq:out_ansatz}
    \Psi(r)=f(r)\phi(z),\hspace{0.5cm}f(r)\equiv(r-2M)^{2iM\omega}\mathrm{e}^{i\omega r},\ \phi(z)\equiv\sum_{n=0}^\infty \phi_n z^n,
\end{equation}
with $z\equiv1-R_{\rm match}/r$, where $R_{\rm match}$ is a radial coordinate located outside the radius of the two-fluid star, which we fix to $R_{\rm match}=1.2R$. The factor $f(r)=(r-2M)^{2iM\omega}\mathrm{e}^{i\omega r}$ in eq.~\eqref{eq:out_ansatz} is the asymptotic form of a purely outgoing wave at spatial infinity when the time dependence is taken as $\mathrm{e}^{i\omega t}$. Indeed, as $r\to\infty$ the tortoise coordinate behaves as $r_*\simeq r+2M\ln[r/(2M)-1]$, so an outgoing wave $\Psi\sim\mathrm{e}^{i\omega r_*}$ reduces to $\Psi\sim(r-2M)^{2iM\omega}\mathrm{e}^{i\omega r}$. By factoring out this leading behavior, the ansatz isolates the residual power series $\phi(z)$, which must be regular at infinity $(z\to1)$.

Introduction of the ansatz into the perturbation equation, rewriting $r$ in terms of $z$ and simplification, yields the following equation for $\phi(z)$:
\begin{equation}
    (a_0+a_1z+a_2z^2+a_3z^3)\frac{d^2\phi}{dz^2}+(b_0+b_1z+b_2z^2)\frac{d\phi}{dz}+(c_0+c_1z)\phi=0,
\end{equation}
where we have defined the dimensionless coefficients:
\begin{align}
    a_0&=1-\frac{2M}{R_{\rm match}},\,a_1=\frac{6M}{R_{\rm match}},\,a_2=1-\frac{6M}{R_{\rm match}},\,a_3=\frac{2M}{R_{\rm match}},\cr
    b_0&=2\left[i\omega R_{\rm match}-1+\frac{3M}{R_{\rm match}}\right],\,b_1=2\left(1-\frac{6M}{R_{\rm match}}\right),\,b_2=\frac{6M}{R_{\rm match}},\cr
    c_0&=\frac{6M}{R_{\rm match}}-\ell(\ell+1),\,c_1=-\frac{6M}{R_{\rm match}}.
\end{align}
Introducing the power series expansion form of $\phi$, we find the recursion relation:
\begin{equation}\label{eq:recursion}
    \begin{cases}
    \alpha_1\phi_2+\beta_1\phi_1+\gamma_1\phi_0=0,\hspace{3.1cm}\text{for }n=1,\cr
    \alpha_n\phi_{n+1}+\beta_n\phi_n+\gamma_n\phi_{n-1}+\delta_n\phi_{n-2}=0,\hspace{0.5cm}\text{for }n\geq2,
    \end{cases}
\end{equation}
with $\alpha_n=n(n+1)a_0$, $\beta_n=nb_0+n(n-1)a_1$, $\gamma_n=c_0+(n-1)b_1+(n-1)(n-2)a_2$ and $\delta_n=c_1+(n-2)b_2+(n-3)(n-2)a_3$. The four-term recurrence relation given in eq.~\eqref{eq:recursion} can be reduced to a three-term recurrence relation by Gaussian elimination~\cite{Leaver:1990zz} of the form:
\begin{equation}
    \tilde{\alpha}_{n}\phi_{n+1}+\tilde{\beta}_{n}\phi_{n}+\tilde{\gamma}_{n}\phi_{n-1}=0,\hspace{0.5cm}\text{for }n\geq1,
\end{equation}
where $\tilde{\alpha}_n=\alpha_n$, $\tilde{\beta}_n=\beta_n-\delta_n\tilde{\alpha}_{n-1}/\tilde{\gamma}_{n-1}$ and $\tilde{\gamma}_n=\gamma_n-\delta_n\tilde{\beta}_{n-1}/\tilde{\gamma}_{n-1}$.
The continued-fraction method consists in defining the ratio of the series coefficients $\mathcal{R}_n\equiv\phi_n/\phi_{n-1}$ and introducing into the recurrence relation to obtain
\begin{equation}\label{eq:cont_frac}
    \mathcal{R}_n=-\frac{\tilde{\gamma}_n}{\tilde{\beta}_n+\tilde{\alpha}_n\mathcal{R}_{n+1}},
\end{equation}
which allows us to obtain $\mathcal{R}_n$ as a function of $\mathcal{R}_{n+1}$. In the case of a convergent series, we have $\mathcal{R}_n\to0$ as $n\to \infty$, so numerically we may fix $\mathcal{R}_{N+1}=0$ for large enough $N$ (we use $N=3000$ in this work) and compute $\mathcal{R}_1=\phi_1/\phi_0$.

On the other hand, from the ansatz equation~\eqref{eq:out_ansatz}, we note that we can write the first two coefficients of $\phi$ as
\begin{align}\label{eq:R_1_inner}
    \phi_0&=\frac{\Psi(R_{\rm match})}{(R_{\rm match}-2M)^{2iM\omega}\mathrm{e}^{i\omega R_{\rm match}}},\cr
    \mathcal{R}_1&=\frac{\phi_1}{\phi_0}=\frac{R_{\rm match}}{\Psi(R_{\rm match})}\left[\left(\frac{d\Psi}{dr}\right)_{r=R_{\rm match}}-\frac{i\omega R_{\rm match}}{R_{\rm match}-2M}\Psi(R_{\rm match})\right].
\end{align}
We note from this equation that we are able to also find the ratio of coefficients $\mathcal{R}_1$ as a function of $\Psi$ and its first derivative evaluated at $r=R_{\rm match}$, which can be obtained from the numerical integration of the inner axial perturbation equation~\eqref{eq:axial_inner}.
The QNMs are then defined as the values of $\omega$ which satisfy that $\mathcal{R}_1$ obtained via the continued-fraction method (eq.~\eqref{eq:cont_frac}) and the one obtained via the inner integration (eq.~\eqref{eq:R_1_inner}) are the same. We have verified that the resulting frequencies are stable under increases of the truncation order $N$ and under moderate variations of the matching radius $R_{\rm match}$. In the numerical solution, we will consider only the case $l=2$.

\section{Quasi-normal modes of boson-fermion stars}\label{sec5}

In this section, we show the results obtained  from the numerical solution, presented in the previous sections, of the axial perturbation. We focus on the behavior of the damping time and frequencies under variation of the central density of each fluid, EoS and the parameter $m_\chi$. 
\subsection{Pure boson star}
Given a QNM written in the form $\omega=\mathrm{Re}[\omega]+i~\mathrm{Im}[\omega]$, we can calculate the damping time as $\tau=|\mathrm{Im}[\omega]|^{-1}$. In a manner analogous to the rescaled variables of eq.~\eqref{eq:rescaled_variables}, we can define the dimensionless QNM and dimensionless damping time as
\begin{equation}
    \omega_*\equiv\omega\frac{\Lambda^{\frac12}}{m_\chi}, \quad \tau_* \equiv |\mathrm{Im}[\omega_*]|^{-1},
\end{equation}
such that we can obtain the general dimensionful expression as a direct function of the microphysical parameters as
\begin{subequations} \label{eq:UnitConversionQNM}
	\begin{align} 
			& \omega=70.21~\mathrm{kHz}~\lambda^{-\frac12}\left(\frac{m_\chi}{1~\rm GeV}\right)^2\omega_*, \label{OmegaConversion} \\
			& \tau=2.27~\mu\mathrm{s}~\lambda^{\frac12}\left(\frac{m_\chi}{1~\rm GeV}\right)^{-2}\tau_*. \label{TauConversion}
	\end{align}
	\end{subequations}
This is a very useful relation, as it shows that we only need to calculate the dimensionless spectrum of QNMs once from which we can obtain the dimensionful one for a given choice of parameters. In fact, we can find empirical expressions for the QNMs. The fundamental mode and the first three overtone modes follow the fits given by the equations below:
\begin{align}\label{eq:BS_empirical}
    &\begin{cases}
    \mathrm{Re}[\omega_{I0,*}]=A\left(\frac{\mathcal{M}_*}{\mathcal{X}^3_*}\right)+B\left(\frac{\mathcal{M}_*}{\mathcal{X}^3_*}\right)^{\frac12}+C,\hspace{1.45cm}\text{for fundamental mode }(i=0),\cr
    \mathrm{Re}[\omega_{Ii,*}]=\frac{1}{\mathcal{X}_*}\left[A\left(\frac{\mathcal{M}_*}{\mathcal{X}_*}\right)^2+B\left(\frac{\mathcal{M}_*}{\mathcal{X}_*}\right)+C\right],\hspace{0.6cm}\text{for }i\geq1
    \end{cases},\cr
    &\tau_{Ii,*}=\frac{1}{\mathcal{M}_*}\left[a\left(\frac{\mathcal{M}_*}{\mathcal{X}_*}\right)^3+b\left(\frac{\mathcal{M}_*}{\mathcal{X}_*}\right)^2+c\left(\frac{\mathcal{M}_*}{\mathcal{X}_*}\right)+d\right].
\end{align}
Here we have defined $\mathcal{X}_*$ as the dimensionless radial coordinate corresponding to the BS's radius and $\mathcal{M}_*\equiv m_*(\mathcal{X}_*)$ to its total dimensionless mass. In table~\ref{table:empirical_fits}, we present the fit parameters $A$, $B$ and $C$ for the dimensionless real part of the first four QNMs ($i\in\{0,1,2,3\}$) and the parameters $a$, $b$, $c$ and $d$ for the corresponding dimensionless damping time.

\begin{table}[h!]
\centering
\small
\renewcommand{\arraystretch}{1.2}
\setlength{\tabcolsep}{3pt}

\caption{Fits for the $\omega_{Ii,*}$ modes. Parameters $A$,  $B$ and $C$ correspond to the empirical relation for the frequency. For $i=0$, we use the first eq. of~\eqref{eq:BS_empirical} and for $i=1,2,3$ we use the second one. Parameters $a$, $b$, $c$ and $d$ correspond to the cubic empirical relation for the dimensionless damping time, also given in eq.~\eqref{eq:BS_empirical}.}
\label{table:empirical_fits}
\begin{tabular}{ccccc}
\hline \hline
$\mathrm{Re}[\omega_{Ii,*}]$ & $i=0$ & $i=1$ & $i=2$ & $i=3$ \\
\hline
$A$	&	$3.906\pm0.006$	&	$-26.2\pm	0.2$ &	$-32.4\pm0.2$ & $-38.6\pm0.2$	 \\
$B$	&	$1.380\pm0.002$	&	$-7.67\pm	0.04$ &	$-15.89\pm0.03$ & $-23.44\pm0.04$	\\
$C$	&	$0.8568\pm0.0001$	&	$5.813\pm0.001$ &	$9.2519\pm0.0008$ & $12.566\pm0.001$	\\
\hline
$\tau_{Ii,*}$ & $i=0$ & $i=1$ & $i=2$ & $i=3$ \\
\hline
$a$	&	$-33.0\pm0.2$	&	$-18.7\pm0.1$ &	$-16.00\pm0.09$ & $-14.44\pm0.09$	\\
$b$	&	$6.56\pm0.05$	&	$2.42\pm0.02$ &	$1.72\pm0.02$ & $1.39\pm0.02$	\\
$c$	&	$1.024\pm0.003$	&	$0.957\pm0.001$ &	$0.878\pm0.001$ & $0.823\pm0.001$	\\
$d$	&	$-0.00072\pm0.00005$	&	$-0.00075\pm0.00002$ &	$-0.00058\pm0.00002$ & $-0.00049\pm0.00002$	\\
\hline
\end{tabular}
\end{table}
\begin{figure}[h!]
	\centering
	\includegraphics[width=.48\textwidth]{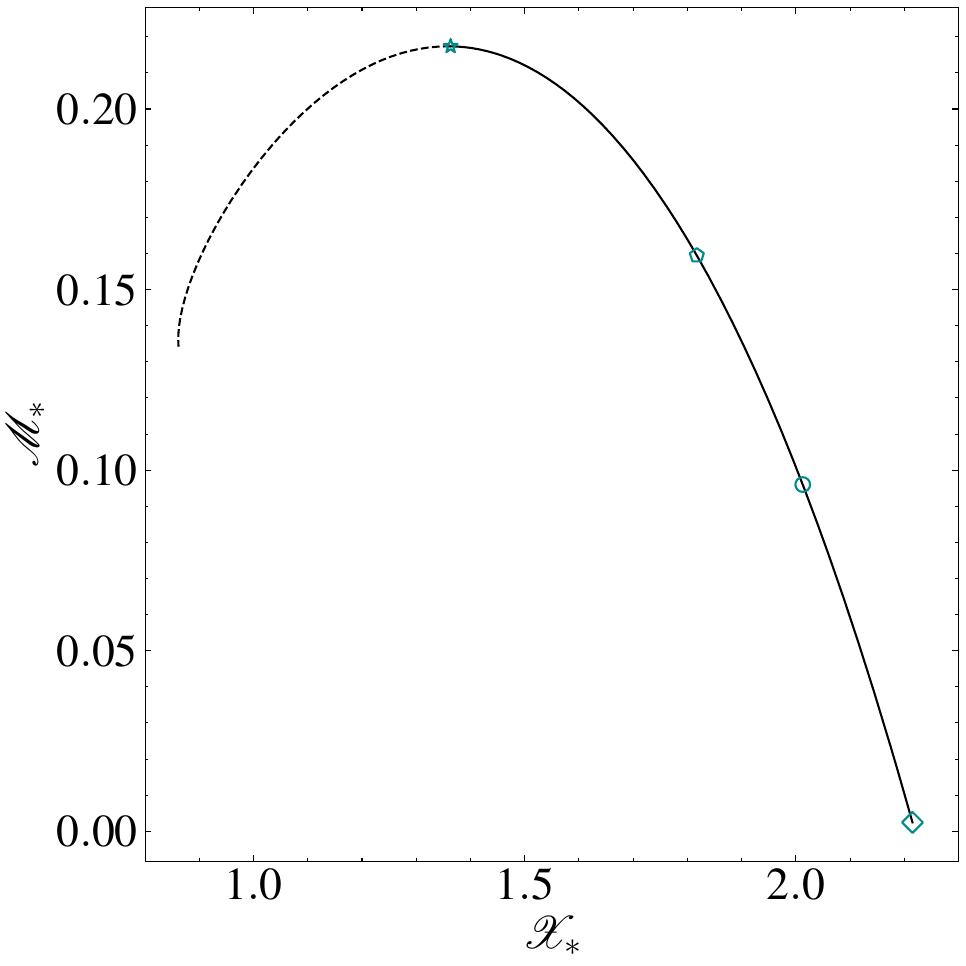}
    \includegraphics[width=.48\textwidth]{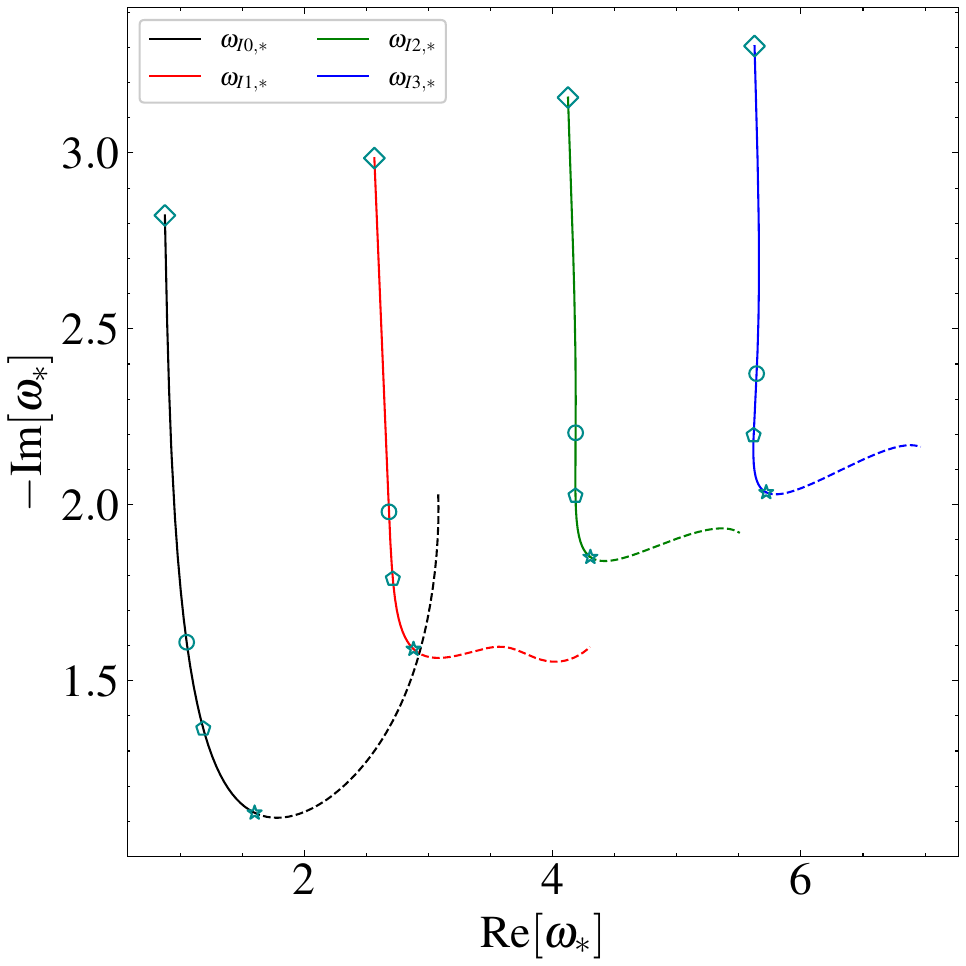}
	\caption{\textit{Left panel}: Dimensionless MR relation for BSs. \textit{Right panel}: Spectrum of dimensionless QNMs. We present the fundamental mode ($\omega_{I0,*}$) and the first three overtones ($\omega_{Ii,*}$, $i\geq1$). In both panels, the solid part of the curve corresponds to solutions that are stable against radial perturbations, while the dashed part is unstable. The scatter points serve as visual cues to relate each macroscopic solution (left panel) with the respective QNM spectrum (right panel).}
	\label{MR_BS}
\end{figure}

\noindent Given that the empirical relations depend on the dimensionless mass and radius, it is advantageous to also provide an empirical fit to the dimensionless MR relation of the BS. It is given by
\begin{align}\label{eq:M_X_dimensionless}
    \mathcal{M}_*=&(-0.04807\pm0.00006)\mathcal{X}_*^3+(-0.0541\pm0.0003)\mathcal{X}_*^2\cr
    &+(0.4115\pm0.0005)\mathcal{X}_*+(-0.1213\pm0.0003).
\end{align}
With both fits given by Eqs.~\eqref{eq:BS_empirical} and~\eqref{eq:M_X_dimensionless} the first four QNMs of the spectrum are completely determined as a function of the dimensionless BS radius. We can observe their behavior graphically in fig.~\ref{MR_BS}. In the left panel of this figure, we present the relation between the dimensionless mass and
radius of the BS. The solid and dashed parts of the curve indicate whether the hydrostatic equilibrium inside the BS is stable or unstable, respectively. Additionally, in the right panel, we show the spectrum of the fundamental and first three overtone QNMs. The points in both panels connect each of the BS solutions with their respective four QNMs. We note that the empirical relations~\eqref{eq:BS_empirical} and~\eqref{eq:M_X_dimensionless} only apply to the stable branches of the BS solutions.

In a similar manner as in eq.~\eqref{eq:Unit_Conversion}, we note that the dimensionful BS QNMs from eq.~\eqref{eq:UnitConversionQNM} are also degenerate in the microphysical parameters. Analogously here, any combination of $\lambda$ and $m_\chi$, such that the product $\lambda\, m_\chi^{-4}$ is the same will yield the same QNMs.

\subsection{Spectrum of quasi-normal modes of boson-fermion stars}

\begin{figure}[h!]
	\centering
	\includegraphics[width=.5\textwidth]{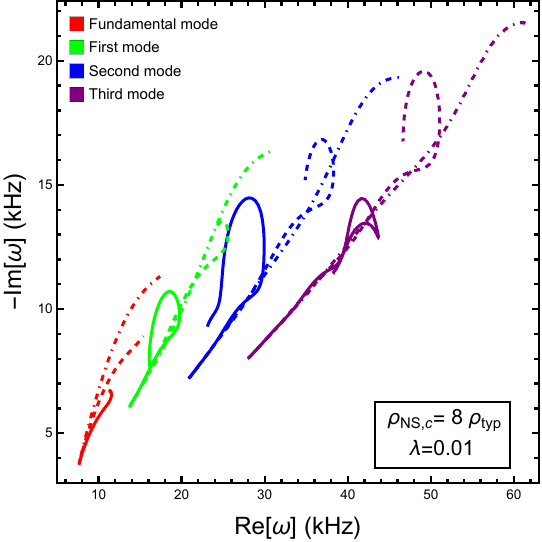}
        
	\caption{First four QNMs  computed using the SLy equation of state by varying the central energy  density $\rho_{\rm BS,c}$  for $m_\chi=0.12$ GeV (dash-dotted curves), $m_\chi=0.10$ GeV (dashed curves) and $m_\chi=0.08$ GeV (solid curves).}
	\label{wrwi}
\end{figure}

\noindent Fig.~\ref{wrwi} presents the  QNM spectrum in the plane $-\mathrm{Im}[\omega]$ versus $\mathrm{Re}[\omega]$, obtained by solving equation (\ref{eq:R_1_inner}), for the fundamental mode and the first three excited axial modes. We consider three different values of $m_\chi$ represented by solid, dashed, and dot-dashed curves. Each color corresponds to a distinct mode and the modes are ordered from the fundamental mode to higher overtones as the real part of the frequency increases. The fundamental mode is located at the lowest frequencies and damping rates, while the first, second, and third excited modes progressively shift toward larger values of both $\mathrm{Re}[\omega]$ and $-\mathrm{Im}[\omega]$. The fundamental mode is the least damped and lowest-frequency oscillation, whereas successive overtones become increasingly short-lived and higher-frequency. To construct this spectrum, we first considered a NS with no bosonic DM with core density $\rho_{\rm NS,c}=8\rho_{\rm typ}$. The first four QNMs of this solution correspond to the approximate values $(\mathrm{Re}[\omega_{Ii}],-\mathrm{Im}[\omega_{Ii}])\approx(8,4),(14,6),(22,7),(28,8)~\mathrm{kHz}$, respectively. As the bosonic DM is introduced and its core density is increased, each QNM shifts depending on the particular choice of $m_\chi$.

The comparison between the solid, dashed, and dot-dashed curves shows that variations of the parameter $m_\chi$ modify the spectrum in a quantitative manner while preserving the same qualitative structure. The three benchmark values of $m_\chi$ have been chosen so that the corresponding pure BS configurations remain comparable to the reference NS solution in both mass and radius: for $m_\chi=0.12~\mathrm{GeV}$ the BS is slightly smaller and less massive ($M_{\rm BS}\sim0.69M_\odot$), for $m_\chi=0.10~\mathrm{GeV}$ it has a mass and radius similar to those of the NS ($M_{\rm BS}\sim1M_\odot$), and for $m_\chi=0.08~\mathrm{GeV}$ it becomes more extended and massive ($M_{\rm BS}\sim1.56M_\odot$). This difference is reflected in the QNM evolution. For $m_\chi=0.12~\mathrm{GeV}$, both $\mathrm{Re}[\omega]$ and $-\mathrm{Im}[\omega]$ generally increase as the bosonic central density is raised, indicating a monotonic shift of the spectrum toward higher frequencies and damping rates. For $m_\chi=0.10~\mathrm{GeV}$, the behavior becomes distinctly non-linear: at low bosonic densities the same increasing trend is observed, but at larger densities the branches may bend back, so that one or both quantities decrease for some modes. This departure from monotonicity is most clearly visible in the second and third overtones. A similar pattern is found for $m_\chi=0.08~\mathrm{GeV}$, although the non-linear behavior is even more pronounced. In general, the higher overtones are more sensitive to these effects than the fundamental mode, which shows that the structure of the axial spectrum carries a stronger imprint of the dark component than the fundamental branch alone.

\subsection{Fundamental axial quasi-normal mode of boson-fermion stars}

In this section, we turn our attention to the fundamental axial mode. In the present two-fluid model, where neutron matter and bosonic  matter are the main components of the star, the axial spectrum is expected to encode the combined influence of both sectors. In this subsection, we investigate the dependence of the fundamental axial mode on the stellar mass, the neutron matter central density $\rho_{\mathrm{NS},c}$, and the bosonic DM central density $\rho_{\mathrm{BS},c}$.

\begin{figure}[h!]
	\centering
	\includegraphics[width=.49\textwidth]{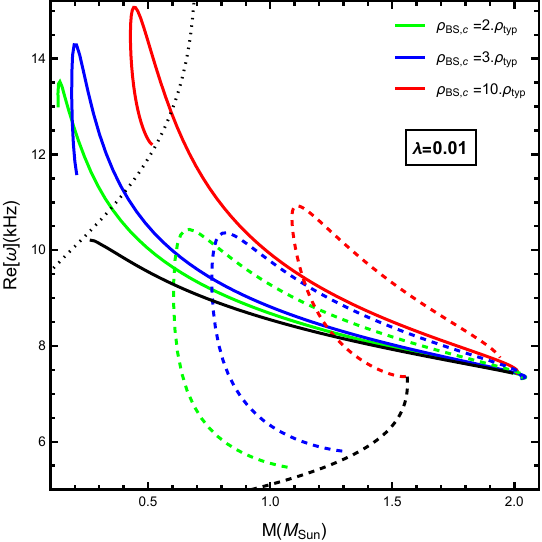}
    \includegraphics[width=.49\textwidth]{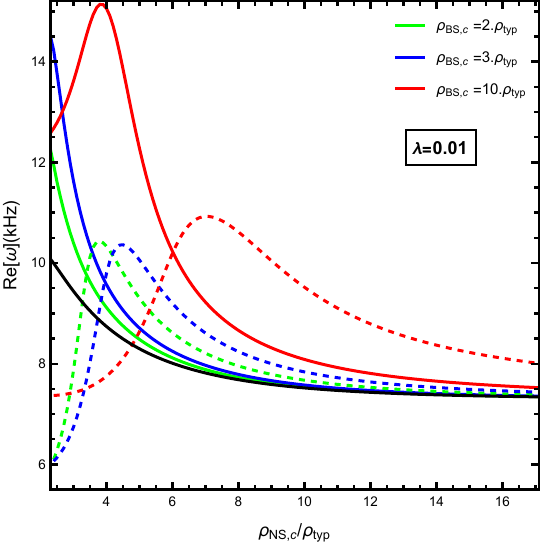} 
	\caption{\textit{Left panel}: Oscillation frequencies as a function of the total mass for different values of the central boson matter energy density, for the SLy equation of state and for the values of $m_\chi=$ $0.12$ GeV (solid lines) and $0.08$ GeV (dashed lines).  Solid black curves correspond to NSs with no DM (matching the results obtained in~\cite{AltahaMotahar:2018djk}) while the dashed and dotted black curves correspond to BSs with no neutron matter for $m_{\chi}=0.08$ GeV and $m_{\chi}=0.12$ GeV, respectively. \textit{Right panel}: Frequencies as a function of $\rho_{\mathrm{NS},c}$ for the same cases shown in the left panel.}
	\label{MRew}
\end{figure}

Fig.~\ref{MRew} presents the real part of the fundamental axial QNM frequency, $\mathrm{Re}[\omega]$, as a function of the total stellar mass $M$ and the neutron-matter central density. The oscillation frequency depends not only on the total mass of the star, but also on the relative contribution of the baryonic and bosonic components. As in the previous section, the deviation from the pure NS case is more pronounced for $m_\chi=0.08~\mathrm{GeV}$, for which the BS component is more massive and more extended. For a fixed total mass, different two-fluid configurations can yield the same oscillation frequency, reflecting the fact that $\mathrm{Re}[\omega]$ is sensitive to the internal matter distribution and not solely to the global mass. In the limit of large $\rho_{\mathrm{NS},c}/\rho_{\mathrm{BS},c}$, the fundamental mode approaches the pure NS value, while for small $\rho_{\mathrm{NS},c}/\rho_{\mathrm{BS},c}$ it tends continuously toward the pure BS branch. This transition is smooth but not monotonic, as the relative central densities are varied, $\mathrm{Re}[\omega]$ may increase or decrease depending on the branch and on the degree of mixing between the two fluids.

\begin{figure}[h!]
	\centering
	\includegraphics[width=.49\textwidth]{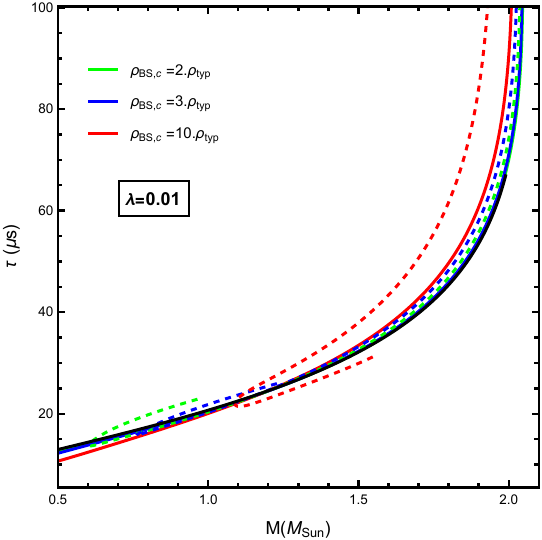}
     \includegraphics[width=.49\textwidth]{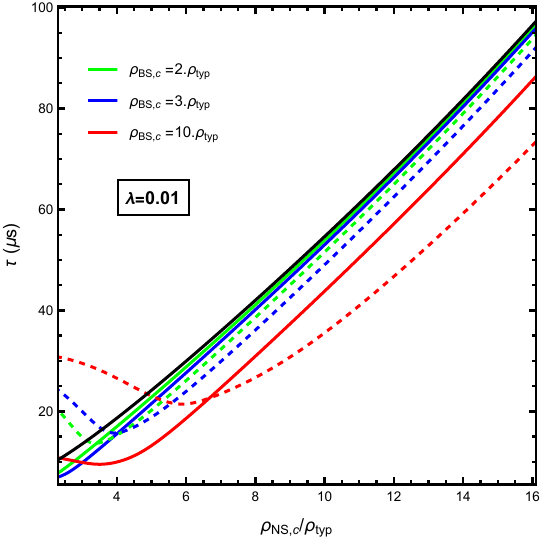}   
	\caption{\textit{Left panel}: Damping time as a function of the total mass for different values of the central boson matter energy density, for the SLy equation of state and for the values $m_\chi=$ $0.12$ GeV (solid lines) and $0.08$ GeV (dashed lines).  Black curves correspond to NSs with no DM (which match the results obtained in~\cite{AltahaMotahar:2018djk}). \textit{Right panel}: Same as left, but as a function of the core NS density.}
	\label{rhodsrhonsw}
\end{figure}

Fig.~\ref{rhodsrhonsw} presents the damping time $\tau$ as a function of the  mass and the neutron matter central density for the same parameters as in fig.~\ref{MRew}. The damping time is related to the imaginary part of the QNM frequency through $\tau = 1/|\mathrm{Im}[\omega]|$ and measures how rapidly the oscillation decays. In general, for a single fluid configuration, larger masses generally have larger damping times. In fact, $\tau$ tends to increase as the star approaches its maximum-mass region. However, in the mixed system with $m_\chi=0.08~\rm GeV$, as in the case of the oscillation frequency, we have two possible damping times for the same total mass. For the case of $\rho_{\rm BS,c}=10\rho_{\rm typ}$, the largest one corresponds to a NS-dominated configuration, while the smallest one to a BS-dominated one. For the other two cases the behavior is mirrored. We show in the right panel of fig.~\ref{rhodsrhonsw}  the variation of  $\tau$ with respect to $\rho_{\mathrm{NS},c}$, where we observe how the damping time of the fundamental axial mode is affected by the baryonic central matter energy density. For low densities, the damping time is more sensitive to increases in the bosonic central energy density and to changes in the bosonic mass $m_\chi$.

\begin{figure}[h!]
	\centering
	\includegraphics[width=.49\textwidth]{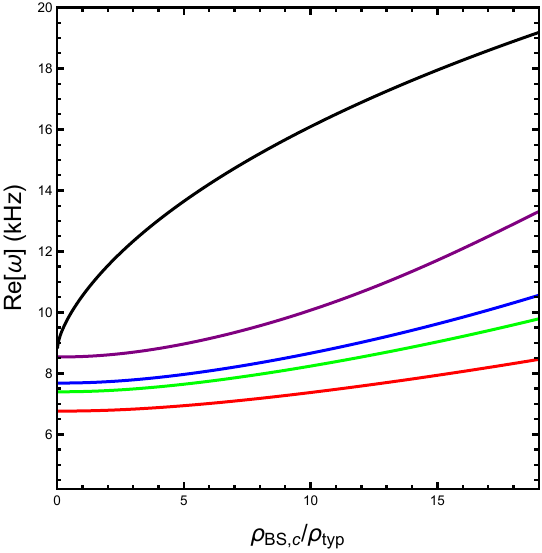}
     \includegraphics[width=.49\textwidth]{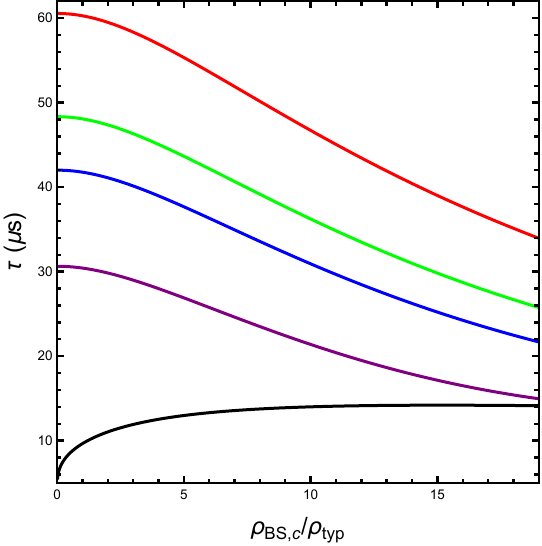}   
	\caption{\textit{Left panel}: Oscillation frequency as a function of the central boson matter energy density for four baryonic EoSs: BSk21 (red), BSk20 (green), SLy (blue) and FPS (purple). \textit{Right panel}: Same as left, but for the damping time. In both panels we fix  $m_\chi=$ $0.12$ GeV, $\rho_{\mathrm{NS},c}=8\rho_{\rm typ}$ and $\lambda=0.01$. In both panels, the black curve corresponds to the pure BS for the same value of $m_\chi$ and $\lambda$.}
	\label{rhodsrhonsw1}
\end{figure}

Fig.~\ref{rhodsrhonsw1} illustrates the dependence of the fundamental axial QNM on the bosonic core density for different baryonic EoSs. For comparison, the black curves show the corresponding oscillation frequency (left panel) and damping time (right panel) for the pure BS configuration. As the bosonic central density increases, the two-fluid configurations evolve continuously from the NS-dominated regime toward the BS limit. In all cases, the oscillation frequency increases while the damping time decreases. Both quantities approach continuously the pure BS behavior independently of the chosen baryonic EoS. Although the qualitative evolution is the same for all EoSs, the rate at which the QNM approaches the BS branch depends on the stiffness of the baryonic sector. The softest EoS considered here, FPS (purple curves), transitions more rapidly toward the BS limit. On the other hand, the stiffest EoS, BSk21 (red curves), remains closer to the pure NS behavior over a larger range of bosonic densities. This behavior can be understood from the different compactness properties associated with each EoS. Softer equations of state produce more compact NS configurations, making the gravitational effect of the bosonic component comparatively more important as the DM fraction increases. By contrast, stiffer EoSs lead to more extended baryonic configurations that remain dynamically relevant even in the presence of a sizeable bosonic core. Consequently, the transition between the NS-like and BS-like QNM occurs at larger bosonic densities for stiffer EoSs.

\begin{figure}[h!]
	\centering
	\includegraphics[width=.49\textwidth]{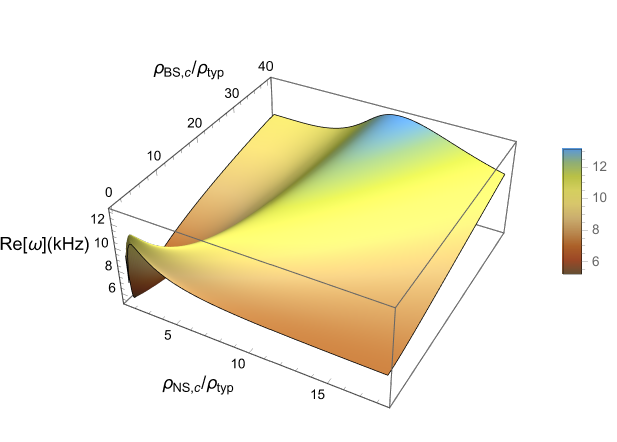}
    \includegraphics[width=.49\textwidth]{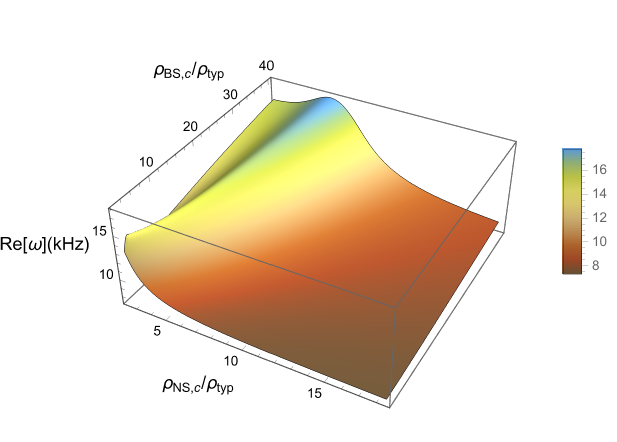}
    \includegraphics[width=.49\textwidth]{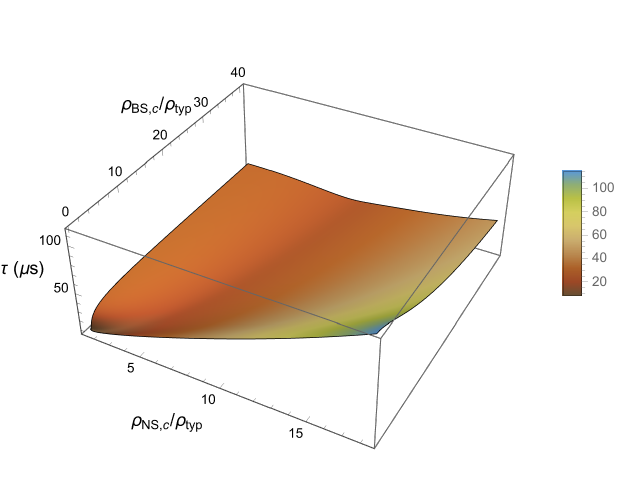}
    \includegraphics[width=.49\textwidth]{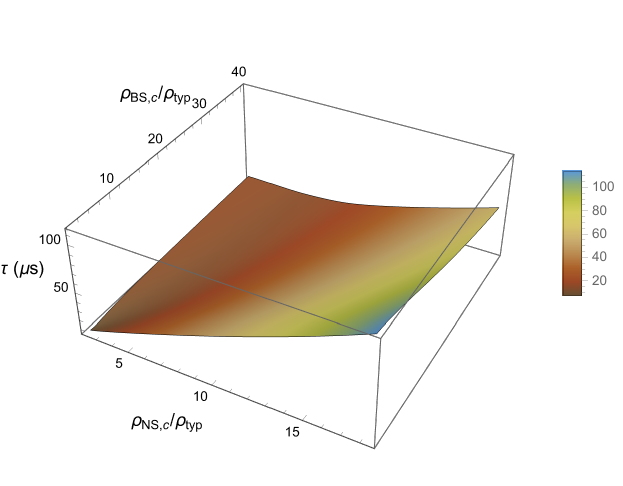}
	\caption{For $\lambda=0.01$ and for the SLy EoS, we show a 3D plot of the oscillation frequency $\mathrm{Re}[\omega]$ and damping time $\tau$ as functions of $\rho_{\rm NS,c}$ and $\rho_{\rm BS,c}$. The left panels correspond to the case $m_{\chi}=0.12~\mathrm{GeV}$ and the right panels correspond to the case $m_{\chi}=0.08~\mathrm{GeV}$.}
	\label{rewrhnsrhobs}
\end{figure}

To illustrate the impact of both central energy densities on the oscillations, we plot in fig.~\ref{rewrhnsrhobs} the three dimensional dependence of the frequency on both central densities, namely $\mathrm{Re}[\omega(\rho_{\mathrm{NS},c},\rho_{\mathrm{BS},c})]$. By increasing $\rho_{\mathrm{BS},c}$, for a fixed $\rho_{\mathrm{NS},c}$, we observe that $\mathrm{Re}[\omega]$ increases for both cases $m_\chi\in\{0.12,0.08\}$ GeV. By contrast, at fixed $\rho_{\mathrm{BS},c}$ the dependence on $\rho_{\mathrm{NS},c}$ is non-monotonic, as the frequency increases up to a maximum and then decreases. This behavior represents the competition between the baryonic and bosonic fluids in setting the effective gravitational potential felt by the axial perturbations and the maximum marks the transition between BS-dominated and NS-dominated. We show also the three-dimensional behavior of the damping time $\tau$ as a function of both the  baryonic central density $\rho_{\rm NS,c}/\rho_{\rm typ}$ and the bosonic central density $\rho_{\rm BS,c}/\rho_{\rm typ}$. The damping time increases with increasing central densities of both fluids, indicating that denser configurations tend to be less damped. In the low density region, the damping time is small compared to larger densities, and we observe that the increase in the damping time is more strongly affected by the baryonic component than by the bosonic component.

In summary, the fundamental axial mode demonstrates that quasi-normal oscillations are highly sensitive to the internal composition of the star. The mode is strongly influenced by the interplay between the baryonic and bosonic components of the star. Its frequency and damping time vary continuously, but not always monotonically, as the central densities of the two fluids are changed, and the deviations from the pure NS case are more pronounced for lighter bosons and for configurations closer to the BS-dominated regime. Although the qualitative behavior is similar for all baryonic EoSs, the approach to the pure BS limit depends on the stiffness of the nuclear EoS.

\subsection{Mode reordering in boson star-dominated configurations}
\begin{figure}[h!]
	\centering
	\includegraphics[width=.48\textwidth]{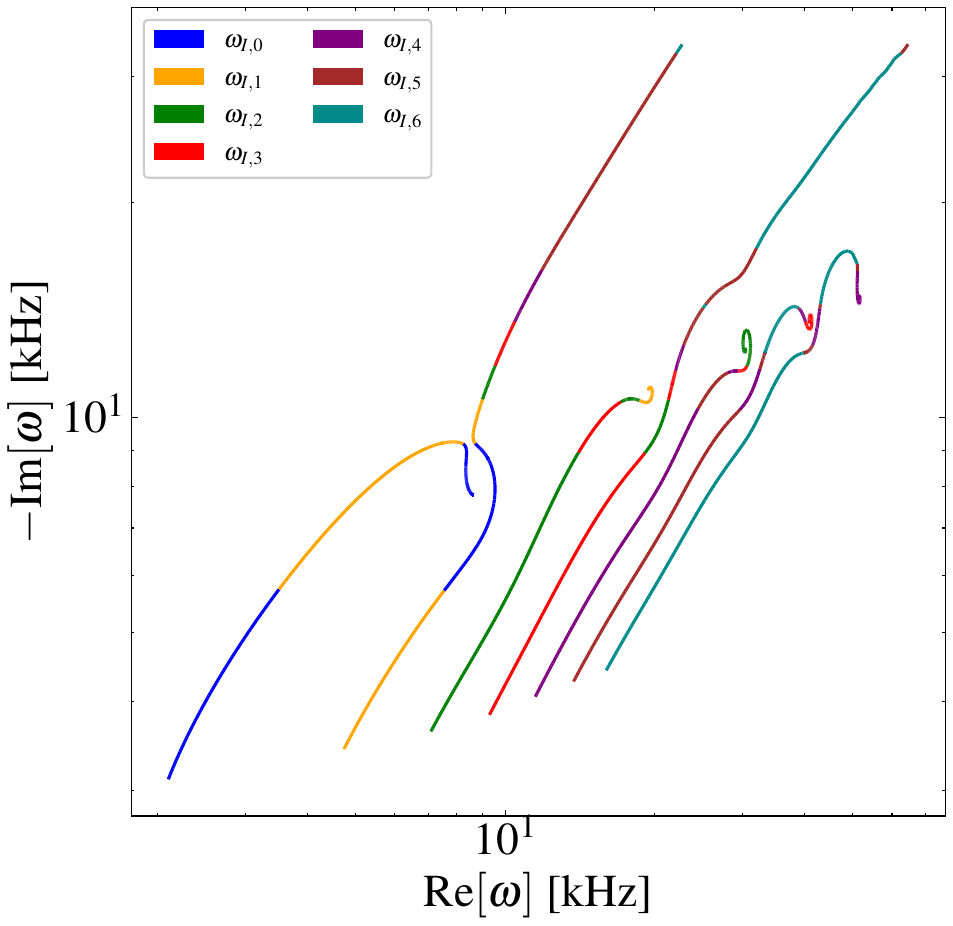}
    \includegraphics[width=.48\textwidth]{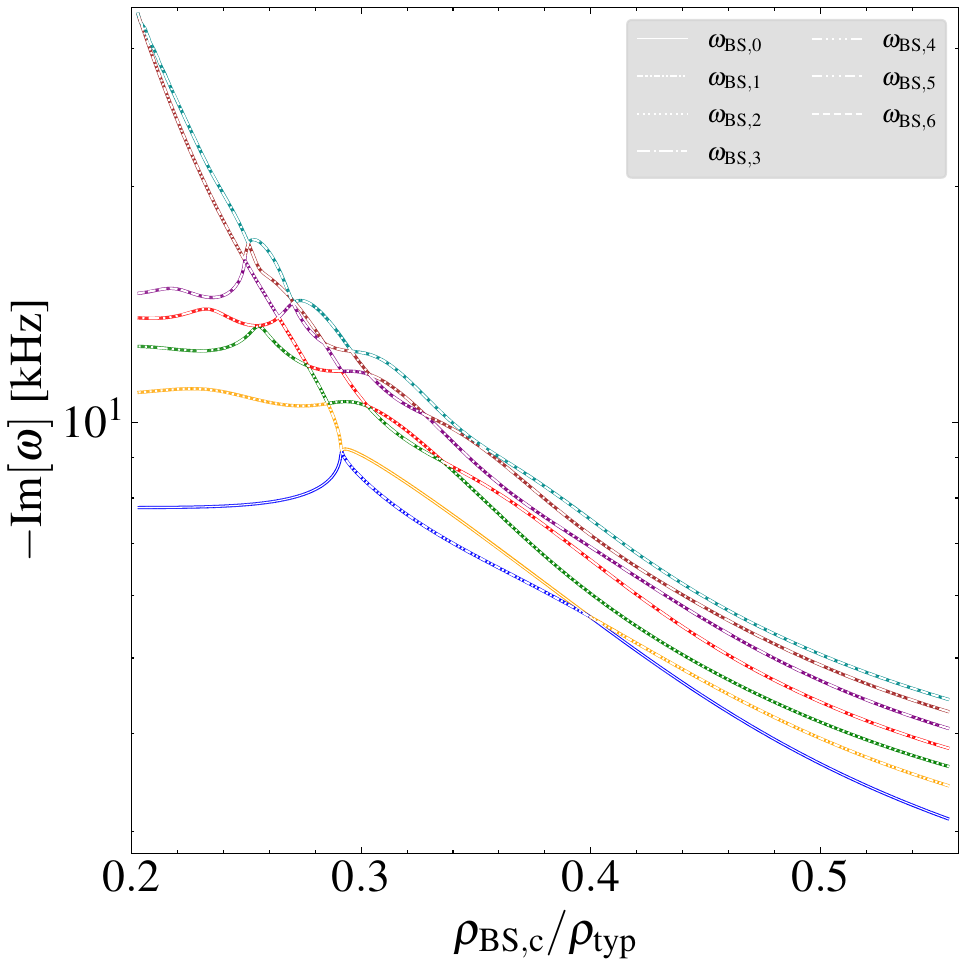}    
	\caption{First seven QNMs computed using the SLy equation of state for fixed $\rho_{\rm NS,c}=4.4\rho_{\rm typ}$ and varying the BS density from $\rho_{\rm BS,c}=0.2\rho_{\rm typ}$ to $\rho_{\rm BS,c}=0.56\rho_{\rm typ}$. The DM parameters are fixed to $\lambda=0.01$ and $m_\chi=0.04~\rm GeV$. \textit{Left panel:} Real and imaginary parts of the QNMs. Each color denotes the magnitude of the imaginary part, with blue corresponding to the smallest and dark cyan to the largest. \textit{Right panel}: Evolution of the imaginary part with the BS central density. The colored curves represent the QNM branches of the boson-fermion star system, maintaining the same color code as the left panel. The white curves trace the same branches backwards from the BS–dominated limit to lower boson densities, explicitly showing how each pure BS mode continuously shifts into (or, in some cases, departs from) the NS spectrum.}
	\label{fig:lowMassSpectrum}
\end{figure}

In this subsection, we consider the case $m_\chi=0.04~\mathrm{GeV}$, for which the isolated BS is more spatially extended and massive ($M_{\rm BS}\sim6.25M_\odot$) than the NS. As a result, the two-fluid configuration does not simply represent a NS with a small DM core, but rather a system that continuously evolves from an NS-like regime to a BS-like one as the bosonic central density increases. This transition is accompanied by a deformation of the axial QNM spectrum, including frequency shifts, mode crossings, and a reordering of the damping hierarchy.

Fig.~\ref{fig:lowMassSpectrum} presents the first seven QNMs of a boson-fermion star configuration, computed using the SLy equation of state for the nuclear matter component. The NS central density is held fixed at $\rho_{\rm NS,c}=4.4\rho_{\rm typ}$, while the bosonic DM content is varied by increasing the BS central density from $\rho_{\rm BS,c}=0.2\rho_{\rm typ}$ to $\rho_{\rm BS,c}=0.56\rho_{\rm typ}$. The DM self-interaction and particle mass are fixed to $\lambda=0.01$ and $m_\chi=0.04~\rm GeV$, respectively. In the left panel, the complex QNM frequencies are shown in the complex plane, with the color scale encoding the magnitude of the imaginary part: blue corresponds to the least damped modes and dark cyan to the most strongly damped ones. The right panel tracks the evolution of the imaginary part of each mode as a function of $\rho_{\rm BS,c}$, which exposes how the QNM spectrum evolves from a NS-dominated regime to a BS-dominated one. The colored curves represent the two-fluid QNM branches and maintain the same color code as the left panel. Superimposed on these are white curves that trace the same branches backwards from the BS-dominated limit toward lower BS densities. Namely, they explicitly show how each pure BS mode continuously shifts into (or, in some cases, disappears from) the NS spectrum.

At the lowest BS density, the DM fraction is small and the spectrum is very close to that of an isolated NS. As the bosonic component increases, the QNM frequencies are shifted away from their pure NS values, showing that the presence of DM produces a continuous deformation of the ringdown spectrum. The shift is initially small and, in fact, from the right panel we note that the magnitude of $\mathrm{Im}[\omega]$ may increase or decrease as the DM density increases depending on the particular mode. As the DM content increases, the deviation becomes increasingly pronounced. Such behavior demonstrates that the boson-fermion star system cannot be viewed as a perturbatively corrected NS alone, rather, the oscillation spectrum reflects the coupled gravitational response of two distinct matter components. In this sense, the QNMs provide a direct probe of how the internal composition of the star changes as the DM fraction is varied. 

An interesting feature shown by the figure is the reordering of the modes as the DM content changes. The fundamental mode, defined as the one with the smallest magnitude of the imaginary part, does not retain this identity throughout the full range of $\rho_{\rm BS,c}$. As the BS density increases, the damping rates of neighboring modes evolve at different rates, and at certain values two modes acquire the same magnitude of the imaginary part. Beyond such a crossing, their ordering is exchanged: the mode that was initially the fundamental becomes the first overtone, while the former first overtone becomes the new fundamental. Such a behavior can be observed via the change in coloring of the spectrum in both panels. We can explicitly see this transition in the right panel at $\rho_{\rm BS,c}/\rho_{\rm typ}\sim0.3$ between the blue and orange curves. This phenomenon is not restricted to the lowest pair of modes. The entire spectrum participates in a cascade of crossings and reorderings, so that modes can move from one position in the damping hierarchy to another multiple times as the bosonic density increases. For example, a mode that starts as the third overtone may successively become the fourth and fifth overtone, and later return to a smaller overtone. The right panel captures this spectral evolution through the intersection of the imaginary-part curves and change of colors, which makes clear that the mode labels are not fixed but rather orderings that depend on the internal composition of the two-fluid system.

As the BS density becomes sufficiently large, the DM component dominates the gravitational potential and the spectrum approaches that of a pure BS. This establishes a continuous transition between the two limiting cases: at low $\rho_{\rm BS,c}$, the system behaves like a NS with a small DM admixture, while at high $\rho_{\rm BS,c}$ the QNM spectrum approaches that of a BS. As mentioned previously, this transition is not just quantitative but also qualitative, since the ordering and character of the QNMs are reorganized along the way. In this regime, the spectrum no longer resembles a small perturbation of the NS case. Instead, it evolves toward the characteristic QNM spectrum of the given BS, which means that the DM core, if dense enough, can ultimately control the dynamical response of the entire object.

The correspondence between the NS and BS spectra is especially revealing when the evolution is followed in the opposite direction, from the BS-dominated regime toward the NS-dominated one. The BS-dominated fundamental mode (denoted by the white, solid line in the right panel) continuously deforms into the NS-dominated fundamental mode as the bosonic density is reduced. Furthermore, in the middle of this evolution, this mode becomes the first overtone of the boson-fermion star system when the DM density is $0.3\lesssim\rho_{\rm BS,c}\lesssim0.4$, denoted by the transition from blue color to orange. But we note that in both the low-DM-density and large-DM-density limits, this mode corresponds to the fundamental one. Nevertheless, the mapping of higher overtones is not one-to-one in the naive sense. For instance, the mode that is the second overtone in the BS-dominated limit does not evolve into the second overtone of the NS; instead, it becomes the first overtone of the NS-dominated spectrum. This shows that the ordering of the branches is significantly reshuffled across the transition, and that simple mode counting does not provide a reliable correspondence between the two limiting spectra. The continuity lies in the branches themselves, but their ranking in the damping hierarchy changes as the relative matter content changes.

An additional feature is the fate of the BS-dominated first and third overtones when the DM density is decreased. In the BS-dominated regime, both modes are well defined, but as $\rho_{\rm BS,c}$ is lowered their imaginary parts increase monotonically and rapidly, indicating progressively stronger damping. Rather than connecting smoothly to one of the NS-dominated QNMs, both branches drift away from the physically relevant part of the spectrum. In that sense, the mode effectively disappears from the long-lived QNM spectrum of the NS-dominated object, namely its damping rate keeps growing, and it does not find a counterpart among the NS modes shown here. This behavior means that some bosonic overtones have a corresponding NS mode across the transition, while others become so damped that they fade out of the observable ringdown window.

The spectral reorganisation shown in fig.~\ref{fig:lowMassSpectrum} raises the question of whether such effects could be detected in gravitational-wave observations and, more importantly, whether they could be distinguished from the changes induced by different neutron-matter EoSs. At the level of a single mode the answer is subtle. For example, for the fundamental $w$I mode shown in fig.~\ref{rhodsrhonsw1}, the frequency at fixed bosonic central density varies significantly across the nuclear EoSs considered, while the shift induced by increasing $\rho_{\rm BS,c}$ at fixed EoS is smaller. The damping time displays a similar behavior. Therefore, a measurement of one axial mode alone would likely remain degenerate with EoS uncertainty.

The situation changes once several overtones are considered simultaneously. The same bosonic component that produces a moderate shift of the fundamental mode also induces mode crossings, a reordering of the damping hierarchy, and the disappearance of some branches into a strongly damped regime, as seen in fig.~\ref{fig:lowMassSpectrum}. A change of the nuclear EoS can modify the overall scale of the spectrum, but it does not naturally reproduce the same branch structure. In this sense, the full QNM pattern carries a more distinctive signature of a two-fluid configuration than any single mode alone.

From an observational point of view, this suggests that EoS degeneracy would have a significant impact on a search based solely on the fundamental $w$ mode. In contrast, as the relative spacing and ordering of the branches depends on both the nuclear EoS and the DM content, a simultaneous measurement of two or more modes would yield a considerably more reliable differentiator. Although it will still be difficult to extract several NS overtones, third-generation detectors, which are anticipated to increase high-frequency sensitivity, may make such an approach more feasible for nearby ringdown sources.

\section{Conclusions}\label{conclusion}

In this work, we have studied the equilibrium structure and axial QNMs of boson-fermion stars within a relativistic two-fluid framework. We first derived the background equations and the EoSs describing the baryonic and dark sectors. For the baryonic matter, we considered the realistic equations of state SLy, FPS, BSk20, and BSk21, while the DM component was modeled as a strongly self-interacting bosonic fluid. We then developed the formalism for axial perturbations and showed that the two-fluid system does not introduce additional dynamical degrees of freedom in the odd-parity sector. Using a continued-fraction method, we computed the complex eigenfrequencies of the associated Regge–Wheeler problem, namely the axial QNMs of boson-fermion stars.

Our results show that the presence of a dark component modifies the macroscopic properties of the compact star and produces a continuous deformation of both the stellar structure and the axial spectrum. As the DM fraction increases, the oscillation frequencies and damping times shift away from their pure NS values, and the changes depend on the boson mass, the self-coupling, the choice of baryonic EoS and the central densities of both fluids.

We found that the fundamental axial mode is particularly sensitive to the internal matter distribution. However, this mode alone does not provide a unique discriminator between DM effects and variations of the nuclear EoS. Consequently, while a single axial mode may reveal deviations from the spectrum of ordinary NSs, it cannot by itself establish the presence of a dark component.

The central result of this work emerges when the bosonic component becomes sufficiently important to drive the system from a NS-dominated regime to a BS-dominated one. In this transition, the QNM spectrum undergoes a reorganization rather than a simple frequency shift. The different branches exchange their ordering in the damping hierarchy through a sequence of level crossings, while some overtones become so strongly damped that they effectively disappear from the observable ringdown. As a consequence, the mode identified as the fundamental oscillation can be replaced by higher overtones, and the spectrum reflects the coupled gravitational response of two distinct matter components rather than a perturbative deformation of the NS case. Such non-trivial spectral reordering constitutes a qualitative signature of a multi-fluid interior that cannot be mimicked by variations of the nuclear equation of state alone, and therefore provides a distinctive imprint of a two-fluid compact object.

 Our results provide the characterization of the axial QNM spectrum of boson-fermion stars within this two-fluid framework and complement recent studies of the polar sector. Although axial modes are not expected to dominate the gravitational-wave signal from astrophysical mergers, they represent the cleanest spacetime sector of the perturbation problem and therefore offer valuable theoretical insight into how an additional self-interacting DM component modifies the oscillation spectrum of compact stars. Future work should extend this analysis to more realistic dynamical scenarios, including tidally deformed stars, rotating configurations, and merger remnants in order to obtain a more complete understanding of the QNM spectrum of exotic compact objects.

 Finally, the present work adopts the relativistic two-fluid description of the bosonic component, which constitutes the appropriate effective description of strongly self-interacting bosonic dark matter in the large self-coupling ($\Lambda>>1$) regime. In this limit, the scalar field is accurately described as an effective perfect fluid, allowing the equilibrium structure and axial perturbations to be treated within a unified two-fluid formalism. Our results therefore characterize the QNM spectrum in this physically well-motivated regime. A complementary extension would be to consider the full Einstein–Klein–Gordon–fluid system, hence relaxing the large-$\Lambda$ approximation and exploring weakly or moderately self-interacting bosonic DM, where qualitatively different equilibrium configurations may arise. Extending the present axial QNM analysis to this broader class of models is a viable direction for future work.

\acknowledgments
The work of BBK is supported by IBS under the project code IBS-R018-D3.

\bibliographystyle{JHEP} 
\bibliography{Bibliography} 

\clearpage

\appendix
\section{The values of the coefficients $a_i$}
In this appendix, we summarize the values of  the coefficients $a_i$ in table~\ref{Tableai}. 

\begin{table}[th!]
\centering \small \addtolength{\tabcolsep}{2pt}
\caption{Parameters of eq.~\eqref{eq:NS_EOS}. SLy and FPS values taken from Ref.~\cite{Haensel:2004nu} and BSk from Ref.~\cite{Potekhin:2013qqa}.}
\begin{tabular}{ccccc}
\hline \hline
$i$ & \multicolumn{4}{c}{$a_i$}         \\
    & SLy & FPS & BSk20 & BSk21 \\
\hline
$1$ & $6.22$    & $6.22$    & $4.078$      & $4.857$      \\
$2$ & $6.121$    & $6.121$    & $7.587$      & $6.981$      \\
$3$ & $0.005925$    & $0.006004$    & $0.00839$      & $0.00706$      \\
$4$ & $0.16326$    & $0.16345$    & $0.21695$      & $0.19351$      \\
$5$ & $6.48$    & $6.50$    & $3.614$     & $4.085$      \\
$6$ & $11.4971$    & $11.8440$    & $11.942$      & $12.065$      \\
$7$ & $19.105$    & $17.24$    & $13.751$      & $10.521$      \\
$8$ & $0.8938$    & $1.065$    & $1.3373$      & $1.5905$      \\
$9$ & $6.54$    & $6.54$    & $3.606$      & $4.104$      \\
$10$ & $11.4950$    & $11.8421$    & $11.942$      & $12.065$      \\
$11$ & $-22.775$    & $-22.003$    & $-22.996$      & $-28.726$      \\
$12$ & $1.5707$    & $1.5552$    & $1.6229$      & $2.0845$      \\
$13$ & $4.3$    & $9.3$    & $4.88$      & $4.89$      \\
$14$ & $14.08$    & $14.19$    & $14.274$      & $14.302$      \\
$15$ & $27.8$    & $23.73$    & $23.560$      & $22.881$      \\
$16$ & $-1.653$    & $-1.508$    & $-1.5564$      & $-1.7690$      \\
$17$ & $1.5$    & $1.79$    & $2.095$      & $0.989$      \\
$18$ & $14.67$    & $15.13$    & $15.294$      & $15.313$      \\
$19$ & $0$    & $0$    & $0.084$      & $0.091$      \\
$20$ & $0$    & $0$    & $6.36$      & $4.68$      \\
$21$ & $0$    & $0$    & $11.67$      & $11.65$      \\
$22$ & $0$    & $0$    & $-0.042$      & $-0.086$      \\
$23$ & $0$    & $0$    & $14.8$      & $10.0$      \\
$24$ & $0$    & $0$    & $14.18$      & $14.15$      \\
\hline
\end{tabular}\label{Tableai}
\end{table}

\end{document}